\documentclass[useAMS,usenatbib]{mn2e}
\usepackage{graphicx,footnote,multirow,booktabs}
\usepackage[T1]{fontenc}
\usepackage{aecompl} 

\title[Transit Timing Analysis in the HAT-P-32 system]{Transit Timing Analysis in the HAT-P-32 system}
\author[M. Seeliger et al.]
{M.~Seeliger,$^{1}$\thanks{E-mail:martin.seeliger@uni-jena.de}
D.~Dimitrov,$^{2}$
D.~Kjurkchieva,$^{3}$
M.~Mallonn,$^{4}$
M.~Fernandez,$^{5}$
\newauthor
M.~Kitze,$^{1}$
V.~Casanova,$^{5}$
G.~Maciejewski,$^{6}$
J.~M.~Ohlert,$^{7,8}$
J.~G.~Schmidt,$^{1}$
\newauthor
A.~Pannicke,$^{1}$
D.~Puchalski,$^{6}$
E.~G\"{o}\u{g}\"{u}\c{s},$^{9}$
T.~G\"{u}ver,$^{10}$
S.~Bilir,$^{10}$
T.~Ak,$^{10}$
\newauthor
M.~M.~Hohle,$^{1}$
T.~O.~B.~Schmidt,$^{1}$
R.~Errmann,$^{1,11}$
E.~Jensen,$^{12}$
D.~Cohen,$^{12}$
\newauthor
L.~Marschall,$^{13}$
G.~Saral,$^{14,15}$
I.~Bernt,$^{4}$
E.~Derman,$^{15}$
C.~Ga{\l}an,$^{6}$
and R.~Neuh\"auser$^{1}$\\
$^{1}~$ Astrophysical Institute and University Observatory Jena, Schillergaesschen 2-3, 07745 Jena, Germany\\
$^{2}~$ Institute of Astronomy and NAO, Bulg. Acad. Sc., 72 Tsarigradsko Chaussee Blvd., 1784 Sofia, Bulgaria\\
$^{3}~$ Shumen University, 115 Universitetska str., 9700 Shumen, Bulgaria\\
$^{4}~$ Leibnitz Institut f\"ur Astrophysik Potsdam, An der Sternwarte 16, 14482 Potsdam, Germany\\
$^{5}~$ Instituto de Astrofisica de Andalucia, CSIC, Apdo. 3004, 18080 Granada, Spain\\
$^{6}~$ Centre for Astronomy, Faculty of Physics, Astronomy and Informatics, N. Copernicus University, Grudziadzka 5, 87-100 Toru\'{n}, Poland\\
$^{7}~$ Astronomie Stiftung Trebur, Michael Adrian Observatorium, Fichtenstra\ss{}e 7, 65468 Trebur, Germany\\
$^{8}~$ University of Applied Sciences, Technische Hochschule Mittelhessen, Friedberg, Germany\\
$^{9}~$ Sabanc\i{} University, Orhanl\i-Tuzla 34956, \.Istanbul, Turkey\\
$^{10}$ Istanbul University, Faculty of Sciences, Department of Astronomy and Space Sciences, 34119 University, Istanbul, Turkey\\
$^{11}$ Abbe Center of Photonis, Friedrich Schiller Universit\"at, Max-Wien-Platz 1, 07743 Jena, Germany\\
$^{12}$ Dept. of Physics and Astronomy, Swarthmore College, Swarthmore, PA 19081-1390, USA\\
$^{13}$ Gettysburg College Observatory, Department of Physics, 300 North Washington St., Gettysburg, PA 17325, USA\\
$^{14}$ Harvard-Smithsonian Center for Astrophysics, 60 Garden Street, Cambridge, MA 02138, USA\\
$^{15}$ Ankara University, Astronomy and Space Sciences Department, 06100 Tando\v{g}an, Ankara, Turkey}

\date{}

\pagerange{\pageref{firstpage}--\pageref{lastpage}} \pubyear{2013}

\begin{document}
\newcommand{\DeltaP}{$21\:$ms}
\newcommand{\DeltaPerror}{$\Delta P=\left(21\pm10\right)\:$ms}
\newcommand{\newP}{$2.15000825$ }
\newcommand{\newPerror}{$0.00000012$}
\newcommand{\newTo}{$2454420.44645$ }
\newcommand{\newToerror}{$0.00009$ }

\newcommand{\newaR}{$6.056$ }
\newcommand{\newaRerror}{$0.009$ }
\newcommand{\newk}{$0.1510$ }
\newcommand{\newkerror}{$0.0004$ }
\newcommand{\newi}{$88.92$ }
\newcommand{\newierror}{$0.10$ }
\newcommand{\Amplitude}{$\sim1.5\:$min}
\newcommand{\Amplitudetwo}{$\sim1.0\:$min}

\label{firstpage}

\maketitle

\begin{abstract}
We present the results of 45 transit observations obtained for the transiting exoplanet HAT-P-32b. The transits have been 
observed using several telescopes mainly throughout the YETI network. In 25 cases, complete transit light curves with a timing 
precision better than $1.4\:$min have been obtained. These light curves have been used to refine the system properties, namely 
inclination $i$, planet-to-star radius ratio $R_\textrm{p}/R_\textrm{s}$, and the ratio between the semimajor 
axis and the stellar radius $a/R_\textrm{s}$. 
First analyses by \citet{Hartman2011} suggest the existence 
of a second planet in the system, thus we tried to find an additional body using the transit timing variation (TTV) technique. 
Taking also literature data points into account, we can explain all mid-transit times by refining 
the linear ephemeris by \DeltaP. Thus we can exclude TTV amplitudes of more than \Amplitude.
\end{abstract}

\begin{keywords}
stars: individual: HAT-P-32 -- planets and satellites: individual: HAT-P-32b -- planetary systems
\end{keywords}

\section{Introduction}
\label{sec:Introduction}
Since the first results of the {\it Kepler} mission were published, the number of known planet candidates has enlarged 
tremendously. Most {\it hot Jupiters} have been found in single planetary systems and it was believed that those kind of 
giant, close-in planets are not accompanied by other planets \citep[see e.g.][]{Steffen2012}. This result has been obtained 
analysing 63 {\it Kepler} hot Jupiter candidates and is in good agreement with inward migration theories of 
massive outer planets, and planet--planet scattering that could explain the lack of additional close planets in hot Jupiter systems. 
Nonetheless, wide companions to hot jupiters have been found, as shown e.g. in \citet{Bakos2009} for the HAT-P-13 system.
One has to state, though, that the formation of hot Jupiters is not yet fully understood (see \citealt{Steffen2012} and 
references therein for some formation scenarios, and e.g. \citealt{Lloyd2013} for possible tests).  
Recently \citet{Szabo2013} reanalysed a larger sample of 159 {\it Kepler} candidates and in some cases found dynamically 
induced {\it Transit Timing Variations (TTVs)}. If the existence of additional planets in hot Jupiter systems 
can be confirmed, planet formation and migration theories can be constrained.

Since, according to \citet{Szabo2013}, there is only a small fraction of hot Jupiters believed to be part of 
a multiplanetary system, it is important to analyse those systems where an additional body is expected.
In contrast to e.g. the Kepler mission, where a fixed field on the sky is monitored over a long time span,
our ongoing study of TTVs in exoplanetary systems only performs follow-up observations of specific promising 
transiting planets where additional bodies are suspected.
The targets are selected by the following criteria: 
\begin{itemize}
	\vspace{-0.4em}
	\setlength{\itemsep}{1pt}
	\setlength{\parskip}{0pt}
	\setlength{\parsep}{0pt}
	\item[(i)] The orbital solution of the known transiting planet shows non-zero eccentricity (though the circularization time-scale 
				is much shorter than the system age) and/or deviant radial velocity (RV) data points -- both indicating a perturber.
	\item[(ii)] The brightness of the host star is $V\leq13\:$mag to ensure sufficient photometric and timing precision at 1-2m telescopes.
	\item[(iii)] The target location on the sky is visible from the Northern hemisphere.
	\item[(iv)] The transit depth is at least 10 mmag to ensure a significant detection at medium-sized, ground-based telescopes.
	\item[(v)] The target has not been studied intensively for TTV signals before.
\end{itemize}

Our observations make use of the YETI network \citep[Young Exoplanet Transit Initiative;][]{YETI}, a worldwide network of small 
to medium sized telescopes mostly on the Northern hemisphere dedicated to explore transiting planets in young open clusters. 
This way, we can observe consecutive transits, which are needed to enhance the possibility to model TTVs as described in \cite{Szabo2013},
and \cite{PTmet}. Furthermore, we are able to obtain simultaneous transits observations to expose hidden systematics in the transit light curves,
like time synchronization errors, or flat fielding errors.

In the past, the transiting exoplanets WASP-12b \citep{Maciejewski2011a, Maciejewski2013a}, WASP-3b \citep{Maciejewski2010,Maciejewski2013b}, 
WASP-10b \citep{Maciejewski2011b,Maciejewski2011c}, WASP-14b \citep{Raetz2012} and TrES-2 (Raetz et al. 2014, submitted) 
have been studied by our group in detail. In most cases, except for WASP-12b, no TTVs could be confirmed. Recently, 
also \citet{vonEssen2013} claimed to have found possible TTV signals around Qatar-1. However, all possible variations should be treated 
with reasonable care.

In this project we monitor the transiting exoplanet HAT-P-32b. The G0V type \citep{Pickles2010} host star HAT-P-32 was found to 
harbour a transiting exoplanet with a period of $P=2.15\:$d by \citet{Hartman2011}. Having a host star brightness of $V=11.3\:$mag 
and a planetary transit depth of $21\:$mmag the sensitivity of medium-sized telescopes is sufficient to achieve high timing 
precision, therefore it is an optimal target for the YETI telescopes. The RV signal of HAT-P-32 is dominated by high 
jitter of $>60\:$ms$^{-1}$.
\citet{Hartman2011} claim that 'a possible cause of the jitter is the presence of one or more additional planets'. 
\citet{Knutson2013} also analysed the RV signature of HAT-P-32 and found a long term trend indicating a companion with a minimum mass
of $5-500\:$M$_{jup}$ at separations of $3.5-12\:$AU. 
However, such a companion could not yet explain the short time-scale jitter as seen in the Hartman data.

Besides the circular orbit fit, an eccentric solution with $e=0.163$ also fits the observed data. 
Though \citet{Hartman2011} mention that the probability of a real non-zero eccentricity is only $\sim3\%$, it could be mimicked or
triggered by a second body in the system. 
Thus, HAT-P-32b is an ideal candidate for further monitoring to look for Transit Timing Variations induced by a planetary companion.


\section{Data acquisition and reduction}
\label{sec:DataAquisitionAndReduction}
Between 2011 October and 2013 January we performed 30 complete and 15 partial transit 
observations (see Tables~\ref{tab:H32completeObservations} and \ref{tab:H32partialObservations}) from 10 different 
observatories: our own one in Jena, as well as from telescopes at Torun (Poland), Trebur (Germany), Gettysburg and 
Swarthmore (USA), Tenerife and Sierra Nevada (Spain), Antalya and Ankara (Turkey), and 
Rozhen (Bulgaria) mostly throughout the YETI network \citep{YETI}. In addition, three literature data points from 
\citet{Sada}, and two observations from \citet{Gibson2013} are available. The telescopes 
and abbreviations used hereafter are summarized in Table~\ref{tab:H32Telescopes}, a short description of each observing 
site can be found below, sorted by the number of observations.
 
\begin{table*}
	\centering
	\caption{The summarized observations gathered within our TTV project for HAT-P-32b listing the telescopes and corresponding 
observatories as well as the telescope diameters $\oslash$ and number of observed transit events (both complete and incomplete) 
in this project N$_{\textrm{tr}}$. At the bottom lines the KPNO telescopes used by \citet{Sada}, and Gemini-North used by 
\citet{Gibson2013} are added for completeness.}
	\label{tab:H32Telescopes}
	
	\begin{tabular}{llllc}
	\toprule
	\# & Observatory                                         & Telescope (abbreviation)                     & $\oslash [m]$ & N$_{tr}$ \\
	\midrule
	1  & University Observatory Jena (Germany)               & Schmidt (Jena 0.6m)                                & 0.6/0.9 & 5 \\
	   &                                                     & Cassegrain (Jena 0.25m)                            & 0.25    & 2 \\
	2  & Teide Observatory, Canarian Islands (Spain)         & STELLA-1 (Tenerife 1.2m)                           & 1.2     & 13\\
	3  & National Astronomical Observatory Rozhen (Bulgaria) & Ritchey-Chr\'{e}tien-Coud\'{e} (Rozhen 2.0m)       & 2.0     & 6 \\
	   &                                                     & Cassegrain (Rozhen 0.6m)                           & 0.6     & 1 \\
	4  & Sierra Nevada Observatory (Spain)                   & Ritchey-Chr\'{e}tien (OSN 1.5m)                    & 1.5     & 7 \\
	5  & Michael Adrian Observatory Trebur (Germany)         & Trebur 1Meter Telescope (Trebur 1.2m)              & 1.2     & 3 \\
	6  & T\"UB\.{I}TAK National Observatory (Turkey)         & T100 (Antalya 1.0m)                                & 1.0     & 1 \\
	7  & Peter van de Kamp Observatory Swarthmore (USA)      & RCOS (Swarthmore 0.6m)                             & 0.6     & 1 \\
	8  & Toru\'n Centre for Astronomy (Poland)               & Cassegrain (Torun 0.6m)                            & 0.6     & 4 \\
	9  & Ankara University Observatory (Turkey)              & Schmidt (Ankara 0.4m)                              & 0.4     & 1 \\
	10 & Gettysburg College Observatory (USA)                & Cassegrain (Gettysburg 0.4m)                       & 0.4     & 1 \\
	11 & Kitt Peak National Observatory (USA)                & 2.1m KPNO Telescope (KPNO 2.1m)                    & 2.1     & 2 \\
	   &                                                     & KPNO Visitor Center Telescope (KPNO 0.5m)          & 0.5     & 1 \\
	12 & Gemini Observatory (Hawaii, USA)                    & Gemini North (GeminiN 8.0m)                        & 8.2     & 2 \\
	\bottomrule
	\end{tabular}

\end{table*}

Most of the observations have been acquired using a defocused telescope. As e.g. \cite{Southworth2009} show, this can be used 
to minimize flat fielding errors, atmospheric effects, and random errors significantly. We ensure that we have enough 
data points acquired during ingress and egress phase to model the transit and get precise transit mid points by having
 at least one data point per minute if possible. The duration of the ingress/egress phase $\tau$ \citep[see][their 
equation~9 assuming zero oblateness]{CarterWinn2010} is given by
\[
	\tau=\left(\frac{R_\textrm{s}\cdot P_{\textrm{orb}}}{\pi\cdot a}\right)\cdot\frac{R_\textrm{p}}{R_\textrm{s}}\cdot\sqrt{1-\left(\cos i\cdot a/R_\textrm{s}\right)^2}\approx24\:\textrm{min}
\]
using the system parameters of \citet{Hartman2011} as listed in Table~\ref{tab:JKTEBOPinput}.
Thus, our observing strategy generates at least 20 data points during ingress/egress phase which allows us to get 
as precise mid-transit times as possible. Using longer exposure times due to e.g. smaller telescope 
sizes has been proven not to improve the fits. Though one can remove atmospheric effects 
better, one loses time resolution at the same time.

The data reduction is performed in a standard way using \textit{IRAF\footnote{IRAF is distributed by the National Optical
Astronomy Observatories, which are operated by the Association of Universities for Research in Astronomy, Inc., under cooperative
agreement with the National Science Foundation.}}. For each respective transit, 
dark or bias frames as well as flat-field images (with the same focus point as the scientific data) in the same 
bands have been obtained during the same night. 

\subsection{Observing telescopes}
\label{sec:ObservingTelescopes}

\noindent \emph{Jena, Germany}

\noindent The University Observatory Jena houses three telescopes. Two of them were used to observe transits of HAT-P-32b. 
The 0.6/0.9m Schmidt Telescope is equipped with an E2V CCD42-10 camera \citep[Jena 0.6m;][]{STK}, and the 0.25m Cassegrain 
Telescope has an E2V CCD47-10 camera (Jena 0.25m). With the first one, we observed three partial and two complete transits, 
the latter one was used for two partial transit observations.

\vspace{\baselineskip}\noindent \emph{Tenerife, Canarian Islands, Spain}

\noindent The robotic telescope STELLA-I, situated at the Teide Observatory and operated by the Leibnitz-Institut 
f\"ur Astrophysik Potsdam (AIP), has a mirror diameter of 1.2m. It is equipped with the Wide Field Stella Imaging Photometer 
\citep[WIFSIP;][]{Weber2012} and could be used to observe seven complete and six partial transit events. The observations 
have been carried out in two filters, $r_\textrm{S}$ and $B$. Since the $r_\textrm{S}$ data turned out to be of higher quality, and to compare 
this data to our other observations, we only used these filter data for further analysis.

\vspace{\baselineskip}\noindent \emph{Rozhen, Bulgaria}

\noindent The telescopes of the National Astronomical Observatory of Rozhen contributed to this study using their 2m telescope 
with a Princ. Instr. VersArray:1300B camera (six complete transit observations), as well as the 0.6m telescope with a FLI 
ProLine 0900 camera (one complete observation).

\vspace{\baselineskip}\noindent \emph{Sierra Nevada, Spain}

\noindent The 1.5m telescope of the Sierra Nevada Observatory, operated by the Instituto de Astrof\'isica de Andaluc\'ia, 
observed seven transits of HAT-P-32b (six complete transit, one partly transit) using a Roper Scientific VersArray 2048B.

\vspace{\baselineskip}\noindent \emph{Trebur, Germany}

\noindent The Trebur One Meter Telescope (T1T, telescope diameter 1.2m) operated at the Michael Adrian Observatory houses 
an SBIG ST-L-6K 3 CCD camera. So far, three complete transits could be observed.

\vspace{\baselineskip}\noindent \emph{Antalya, Turkey}

\noindent The T100 Telescope of the T\"UB\.{I}TAK National Observatory observed one transit of HAT-P-32b using a Spectral 
Instruments 1100 series CCD camera. However, due to technical problems the observation had to be cancelled.

\vspace{\baselineskip}\noindent \emph{Swarthmore, USA}

\noindent The 0.6m telescope at the Peter van de Kamp Observatory of Swarthmore College contributed one complete transit 
observation using an Apogee U16M KAF-16803 CCD camera.

\vspace{\baselineskip}\noindent \emph{Toru\'{n}, Poland}

\noindent One partial and three complete transits have been observed using the 0.6m Cassegrain Telescope at the Toru\'{n} 
Centre for Astronomy with an SBIG ST-L-1001 CCD camera mounted.

\vspace{\baselineskip}\noindent\emph{Ankara, Turkey}

\noindent On 2011-10-04 a complete transit of HAT-P-32b was observed using the 0.4m Schmidt-Cassegrain Meade LX200 GPS 
Telescope equipped with an Apogee ALTA U47 CCD camera located at and operated by the University of Ankara.

\vspace{\baselineskip}\noindent \emph{Gettysburg, USA}

\noindent The 0.4m Cassegrain reflector from Gettysburg College Observatory was used to observe one complete transit on 
2012-01-15. Data were obtained with the mounted CH350 CCD camera with a back-illuminated SiTE3b chip with an $R$-filter.

\begin{table}
\begin{minipage}{1\columnwidth}
	\centering
	\caption{The list of complete and usable transit observations gathered within the TTV project for HAT-P-32b. The initial 
ephemeris from the discovery paper \citep{Hartman2011}, and three data points from the literature obtained at KPNO \citep{Sada} 
are given at the bottom lines. For the latter ones, no exposure times are given. Filter indices B, C, and S denote the photometric 
systems Bessel, Cousins, and Sloan, respectively, used with the different instrumentations.} \label{tab:H32completeObservations}
	\begin{tabular}{llllr}
	\toprule
	\# & epoch\footnote{The epoch is calculated from the originally published ephemeris by \cite{Hartman2011}.} & Telescope                                  & filter & exposure $[s]$ \\
	\midrule
	1  &  673  & Tenerife 1.2m                              & $r_\textrm{S}$ & 15            \\
	2  &  679  & Jena 0.6m                                  & $R_\textrm{B}$ & 40            \\
	3  &  686  & Rozhen 2.0m                                & $V_\textrm{C}$ & 20            \\
	4  &  686  & Tenerife 1.2m                              & $r_\textrm{S}$ & 15            \\
	5  &  687  & Rozhen 2.0m                                & $R_\textrm{C}$ & 20            \\
	6  &  693  & Rozhen 0.6m                                & $R_\textrm{C}$ & 60            \\
	7  &  699  & Rozhen 2.0m                                & $R_\textrm{C}$ & 20            \\
	8  &  708  & Swarthmore 0.6m                            & $R_\textrm{C}$ & 50            \\
	9  &  807  & Rozhen 2.0m                                & $R_\textrm{C}$ & 25            \\
	10 &  808  & Tenerife 1.2m                              & $r_\textrm{S}$ & 15            \\
	11 &  820  & OSN 1.5m                                   & $R_\textrm{C}$ & 30            \\
	12 &  820  & Trebur 1.2m                                & $R_\textrm{B}$ & 50            \\
	13 &  821  & OSN 1.5m                                   & $R_\textrm{C}$ & 30            \\
	14 &  833  & Trebur 1.2m                                & $R_\textrm{B}$ & 50            \\
	15 &  853  & OSN 1.5m                                   & $R_\textrm{C}$ & 30            \\
	16 &  873  & Tenerife 1.2m                              & $r_\textrm{S}$ & 25            \\
	17 &  987  & Jena 0.6m                                  & $R_\textrm{B}$ & 40            \\
	18 &  987  & Rozhen 2.0m                                & $R_\textrm{C}$ & 30            \\
	19 &  987  & Torun 0.6m                                 & $clear       $ & 10            \\
	20 & 1001  & OSN 1.5m                                   & $R_\textrm{C}$ & 30            \\
	21 & 1013  & Rozhen 2.0m                                & $R_\textrm{C}$ & 25            \\
	22 & 1014  & OSN 1.5m                                   & $R_\textrm{C}$ & 30            \\
	23 & 1027  & OSN 1.5m                                   & $R_\textrm{C}$ & 30            \\
	24 & 1040  & Trebur 1.2m                                & $R_\textrm{B}$ & 50            \\
	\midrule
	25 &  \phantom{00}0 & \multicolumn{3}{l}{see \citet{Hartman2011} }          \\
	26 &  662  & \multicolumn{3}{l}{KPNO 2.1m; \citet{Sada}           }          \\
	27 &  663  & \multicolumn{3}{l}{KPNO 2.1m; \citet{Sada}          }           \\
	28 &  663  & \multicolumn{3}{l}{KPNO 0.5m; \citet{Sada}          }           \\
	\bottomrule
	\end{tabular}
\end{minipage}
\end{table}

\begin{table*}\begin{minipage}{1\textwidth}
	\centering
	\caption{The list of partial transit observations or unusable observations gathered within the TTV project for HAT-P-32b.}\label{tab:H32partialObservations}
	
	\begin{tabular}{llllrl}
	\toprule
	\# & epoch\footnote{The epoch is calculated from the originally published ephemeris by \cite{Hartman2011}.} & Telescope&filter & exposure $[s]$& remarks\\
	\midrule
	1  &  654  & Jena 0.6m                                    & $R_\textrm{B}$ & 40  & only first half of transit observed\\
	2  &  660  & Ankara 0.4m                                  & $R_\textrm{C}$ & 10  & large fit errors              \\
	3  &  666  & Torun 0.6m                                   & $R$            & 30  & bad observing conditions\\
	4  &  673  & Jena 0.6m                                    & $R_\textrm{B}$ & 50  & only first half of transit observed\\
	5  &  693  & Tenerife 1.2m                                & $r_\textrm{S}$ & 15  & only ingress observed         \\
	6  &  700  & Tenerife 1.2m                                & $r_\textrm{S}$ & 15  & only second half of transit observed\\
	7  &  708  & Gettysburg 0.4m                              & $R$            & 50  & bad observing conditions\\
	8  &  713  & Jena 0.6m                                    & $R_\textrm{B}$ & 50  & only first half of transit observed\\
	9  &  807  & Antalya 1.0m                                 & $R$            & 3   & technical problems\\
	10 &  821  & Tenerife 1.2m                                & $r_\textrm{S}$ & 10  & bad weather during egress phase         \\
	11 &  833  & Jena 0.25m                                   & $R_\textrm{B}$ & 100 & bad weather, gaps in the data\\
	12 &  834  & Jena 0.25m                                   & $R_\textrm{B}$ & 100 & only first half of transit observed\\
	13 &  834  & OSN 1.5m                                     & $R_\textrm{C}$ & 20  & upcoming bad weather during ingress\\
	14 &  840  & Tenerife 1.2m                                & $r_\textrm{S}$ & 10  & large fit errors   \\
	15 &  848  & Tenerife 1.2m                                & $r_\textrm{S}$ & 15  & only egress phase observed         \\
	16 &  854  & Tenerife 1.2m                                & $r_\textrm{S}$ & 20  & only ingress phase observed         \\
	17 &  855  & Tenerife 1.2m                                & $r_\textrm{S}$ & 20  & jumps in data, no good fits possible         \\
	18 &  861  & Tenerife 1.2m                                & $r_\textrm{S}$ & 25  & bad observing conditions         \\
	19 &  867  & Tenerife 1.2m                                & $r_\textrm{S}$ & 25  & bad observing conditions          \\
	20 &  906  & Torun 0.6m                                   & $clear$        & 10  & only ingress phase observed\\ 
	21 & 1001  & Torun 0.6m                                   & $clear$        & 6   & jumps in data, no good fits possible  \\
	\bottomrule
	\end{tabular}\end{minipage}
\end{table*}

\section{Analysis}
\label{sec:Analysis}

All light curves (except for the Ankara 0.4m observation) are extracted from the reduced images using the same 
aperture photometry routines (described in section~\ref{sec:GettingTheLightcurve}) 
in order to prevent systematic offsets between different transit observations due to different light curve 
extraction methods. Afterwards we model the data sets using the \textit{JKTEBOP} 
algorithm \citep{JKTEBOP}, as well as the Transit Analysing Package \textit{TAP} \citep{TAP} as described in the following sections.

\subsection{Obtaining the light curve}
\label{sec:GettingTheLightcurve}

Before generating the light curve, we compute the Julian Date of each exposure midtime using the header informations of each image. 
Hence, the quality of the final light curve fitting is not only dependent on the photometric precision, but also on a precise time 
synchronization of the telescope computers. One good method to reveal synchronization problems is to observe one transit from 
different telescope sights as done at epochs 673, 686, 693, 708, 807, 820, 821, 833, 834, and 987 (see Tables~\ref{tab:H32completeObservations} 
and \ref{tab:H32partialObservations}). Unfortunately, due to bad weather conditions, some of the observations had to be aborted 
or rejected after a visual inspection of the light curve.

The brightness measurements are done with \textit{IRAF} performing aperture photometry on all bright stars in 
the field of view in each image obtained per transit. The aperture size is manually varied to find the best photometric precision. 
Typical aperture values are $\sim$1.5 times the full width half-maximum. To generate the transit light curve (including the error), 
we use differential aperture photometry by computing a constant artificial standard star containing all comparison stars with a 
brightness of up to 1~mag fainter than the target star as introduced by \citet*{Broeg2005}.
The weight of each star is computed by its constancy and photometric precision, thus including more fainter stars does not increase 
the precision of the artificial star and hence the final light curve. 
The photometric error of each data point is rescaled using the \textit{IRAF} measurement error, and the standard deviation in the 
light curves of all comparison stars as scaling parameters (for a more detailed description, see \citealt{Broeg2005}).

Due to the small field of view, the Ankara 0.4m observation has been treated different. After applying the standard 
image reduction, differential aperture photometry was used to create the light curve. As comparison star, we used GSC~3280~781, i.e. 
the brightest, unsaturated star in the field of view.

To prepare the final light curve to be modelled, we fit a quadratic trend (second order polynomial) to the normal light phases to 
adjust for secondary airmass effects. Thus, it is required and ensured to observe a complete transit with one hour of normal light 
before and after the transit event itself. In addition to the originally data, we also binned all light curves threefold using an error 
weighted mean.

As a quality marker for our light curves we derive the photometric noise rate ($pnr$) as introduced by \citet{Fulton2011}. 
The $pnr$ is calculated using the root mean square ($rms$) of the model fit, as well as the number of data points per minute $\Gamma$. 
\begin{center}$pnr=\frac{rms}{\sqrt{\Gamma}}$\end{center}
The respective values for all our modelled light curves are given in Table~\ref{tab:fitResults} together with the fitted system parameters.

\subsection{Modelling the light curve with JKTEBOP}
\label{sec:JKTEBOP}

The light curve model code \textit{JKTEBOP} \citep[see e.g.][]{JKTEBOP}, based on the \textit{EBOP} code \citep{Etzel1981,Popper1981}, 
fits a theoretical light curve to the data using the parameters listed in Table~\ref{tab:JKTEBOPinput}. 
Since we only deal with ground based data, we only take quadratic limb darkening (LD) into account to directly 
compare the results of the fitting procedure with those of \textit{TAP} (see the next section). We employ the LD 
values from \cite{Claret2000} for the stellar parameters listed in Table~\ref{tab:JKTEBOPinput}. To get the values we use the \textit{JKTLD} 
code\footnotemark\footnotetext{see http://www.astro.keele.ac.uk/jkt/codes/jktld.html} that linearly interpolates between the model grid 
values of $T_{eff}$ and $\log g$. Since \citet{Hartman2011} only list $\left[ Fe/H\right]$ instead of $\left[ M/H\right]$ that is needed 
to get the LD values, we converted it according to \citet{Salaris1993}. \textit{JKTLD} does not interpolate 
for $\left[ M/H\right]$, hence a zero value was assumed, which is consistent with the known values within the error bars. Since most
LD coefficients are only tabulated for $V_{micro}$ = 2 km s$^{-1}$, this value was adopted (as also suggested 
by J. Southworth\footnotemark[2]).

\begin{table}
	\centering
	
	\caption{The input parameters for the JKTEBOP \& TAP runs with values of the circular orbit fit from the discovery paper 
\citep{Hartman2011}, as well as the derived limb darkening coefficients. The metalicity of $\left[ Fe/H\right]=\left(-0.04\pm0.08\right)\,$dex 
according to the circular orbit fit of \citet{Hartman2011} was converted to $\left[ M/H\right]=\left(-0.03\pm0.11\right)\,$dex using the 
equations of \citet{Salaris1993}. Free-to-fit parameters are marked by an asterisk.}\label{tab:JKTEBOPinput}
	{\setlength{\tabcolsep}{4pt}
	\begin{tabular}{lr@{$\,\pm\,$}l}
		\toprule
		parameter & \multicolumn{2}{c}{value}\\
		\midrule
		sum of radii $r_\textrm{p}+r_\textrm{s}$*				&	 0.1902&0.0013\\
		ratio of radii $R_\textrm{p}/R_\textrm{s}$*				&	 0.1508 &0.0004\\
		orbital inclination $i$ [$^\circ$]*						&	88.9 & 0.4  \\
		inverse fractional stellar radius $a/R_\textrm{s}$*		&	 6.05 &0.04 \\
		mass ratio of the system $M_\textrm{p}/M_\textrm{s}$	&	 0.0007 &  0.0002\\
		orbital eccentricity $e$								&	\multicolumn{2}{c}{ 0}		\\
		orbital period $P$ [d]									&	 2.150008 & 0.000001\\
		\midrule
		T$_{\textrm{eff}}$ [K]									&	6207 & 88  \\
		$\log g$ [cgs]											&	4.33 & 0.01 \\
		$[Fe/H]$ [dex]											&	-0.04&  0.08 \\
		$[M/H]$ [dex]											&	-0.03 & 0.11 \\
		$v \sin i$ [km s$^{-1}$]								&	20.7 & 0.5  \\
		$V_{micro}$[km s$^{-1}$]								& \multicolumn{2}{c}{2}\\
		\midrule
		limb darkening (LD) law of the star						& \multicolumn{2}{c}{quadratic}\\
		linear LD coefficient R-band*							&	\multicolumn{2}{c}{0.28}	\\
		nonlinear LD coefficient R-band*						&	\multicolumn{2}{c}{0.35}	\\
		linear LD coefficient V-band*							&	\multicolumn{2}{c}{0.37}	\\
		nonlinear LD coefficient V-band*						&	\multicolumn{2}{c}{0.34}	\\
		\bottomrule
	\end{tabular}
	}
\end{table}

We fit for the parameters mid-transit time $T_{\textrm{mid}}$, sum of the fractional radii $r_\textrm{p}+r_\textrm{s}$ 
($r_\textrm{p}$ and $r_\textrm{s}$ being the radius of the planet $R_\textrm{p}$ and the star $R_\textrm{s}$ divided by the semimajor 
axis $a$, respectively), ratio of the radii $R_\textrm{p}/R_\textrm{s}$, and orbital inclination $i$. In case of the limb darkening 
coefficients we use two different configurations having the linear and nonlinear term fixed, and fitting them around the theoretical 
values, respectively. The eccentricity is assumed to be zero.

\textit{JKTEBOP} allows us to apply different methods to estimate error bars. We used Monte Carlo simulations (10$^4$ runs), bootstrapping 
algorithms (10$^4$ data sets), and a residual-shift method to see if there are significant differences in the individual error estimation methods\footnotemark.
\footnotetext{see http://www.astro.keele.ac.uk/jkt/codes/jktebop.html and references therein for details about the code and error estimation methods}

\subsection{Modelling the light curve with TAP}
\label{sec:TAP}

The Transit Analysis Package \textit{TAP}\footnotemark\footnotetext{http://ifa.hawaii.edu/users/zgazak/IfA/TAP.html} \citep{TAP} 
makes use of the EXOFAST routine \citep{Eastman2013} with the light curve model by \cite{MandelAgol2002} and the wavelet-based 
likelihood functions by \cite{Carter2009} to model a transit light curve and to estimate error bars. The input parameters are 
listed in Table~\ref{tab:JKTEBOPinput}. \textit{TAP} only uses quadratic limb darkening. Instead of the sum of fractional radii 
$r_\textrm{p}+r_\textrm{s}$, that is used by \textit{JKTEBOP}, \textit{TAP} uses the inverse fractional stellar radius $a/R_\textrm{s}$,
but those two quantities can be converted into each other using the ratio of radii.

\begin{equation}
	a/R_\textrm{s}=\left(1+R_\textrm{p}/R_\textrm{s}\right)/\left(r_\textrm{p}+r_\textrm{s}\right)
\end{equation}
With \textit{TAP} we also model the light curve several times using the unbinned and binned data (see section~\ref{sec:GettingTheLightcurve}), 
as well as keeping the limb-darkening coefficients fixed and letting them vary around the theoretical values. To estimate error bars, 
\textit{TAP} runs several (in our case 10) Markov Chain Monte Carlo (MCMC) simulations with 10$^5$ steps each.

\section{Results}
\label{sec:Results}

\begin{figure*}
  \includegraphics[width=0.29\textwidth]{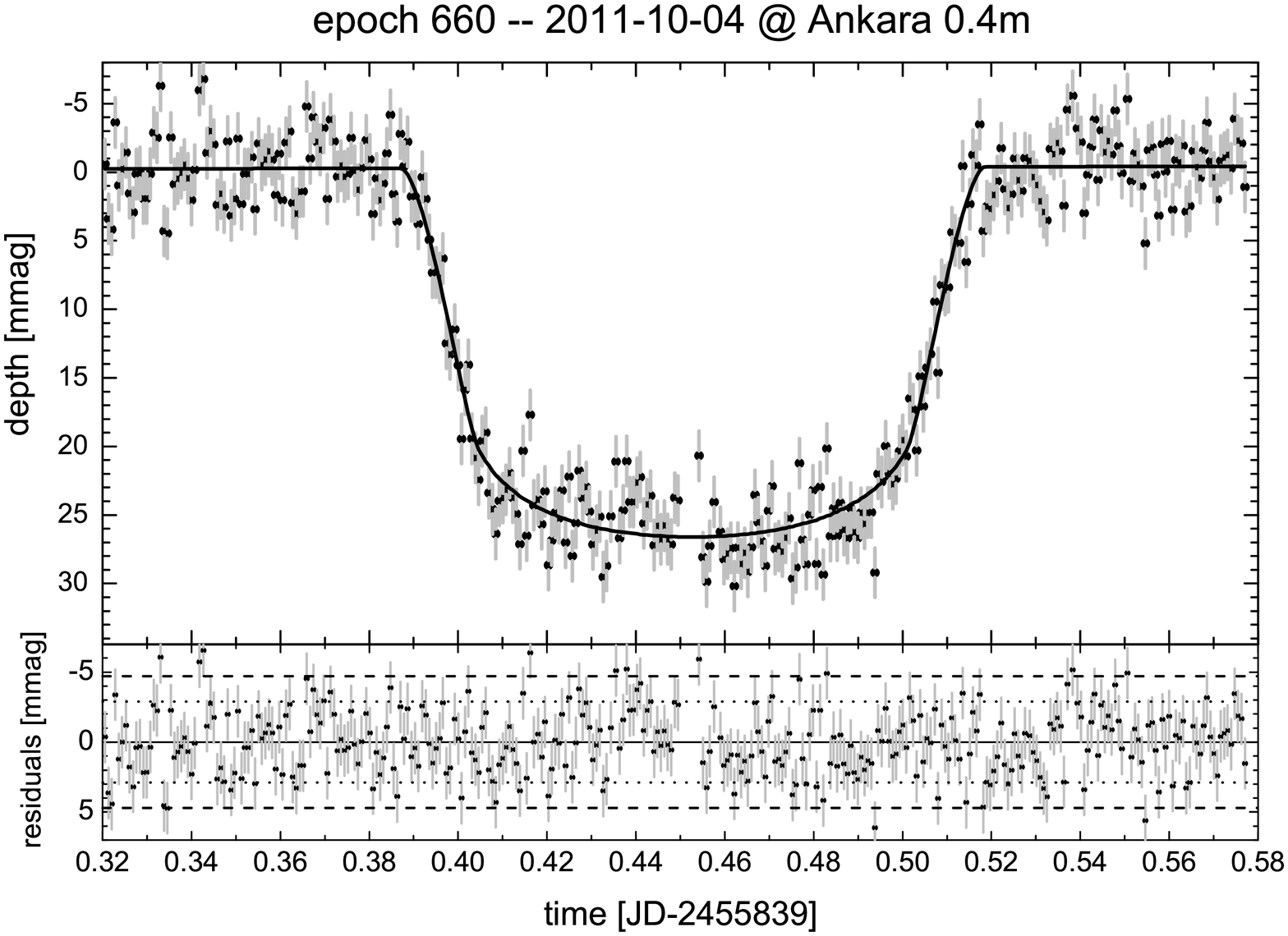}
  \includegraphics[width=0.29\textwidth]{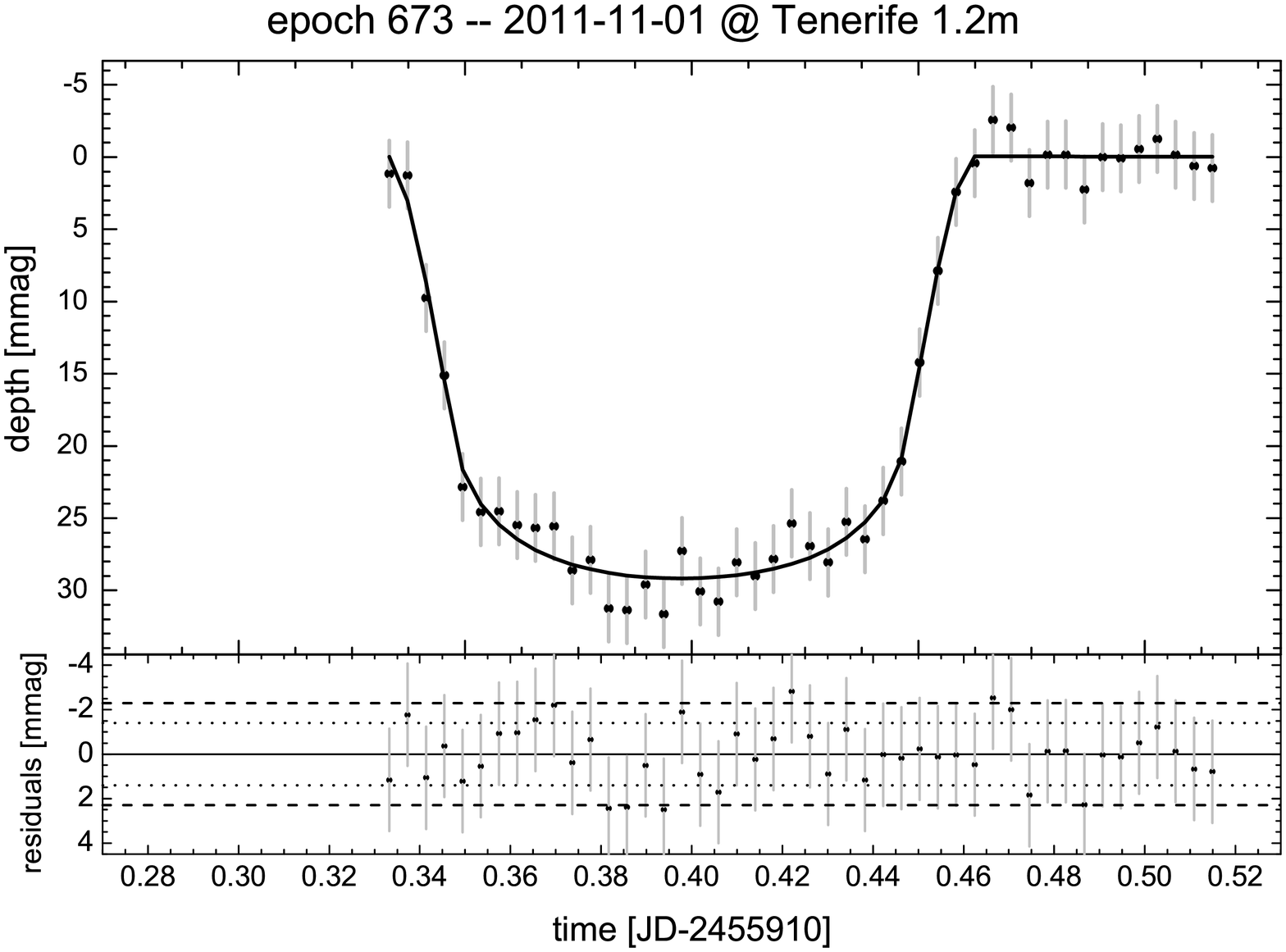}
  \includegraphics[width=0.29\textwidth]{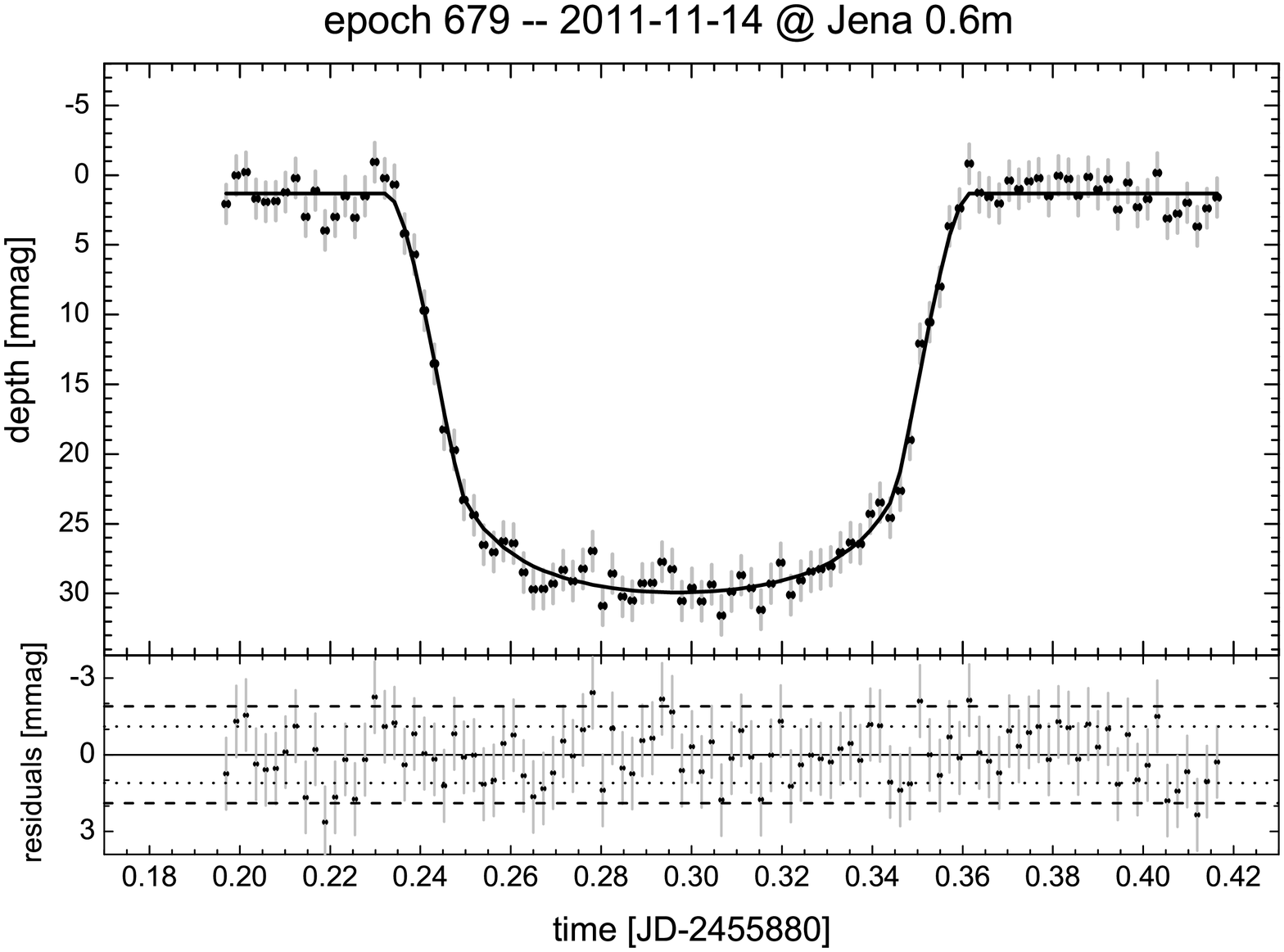}

  \includegraphics[width=0.29\textwidth]{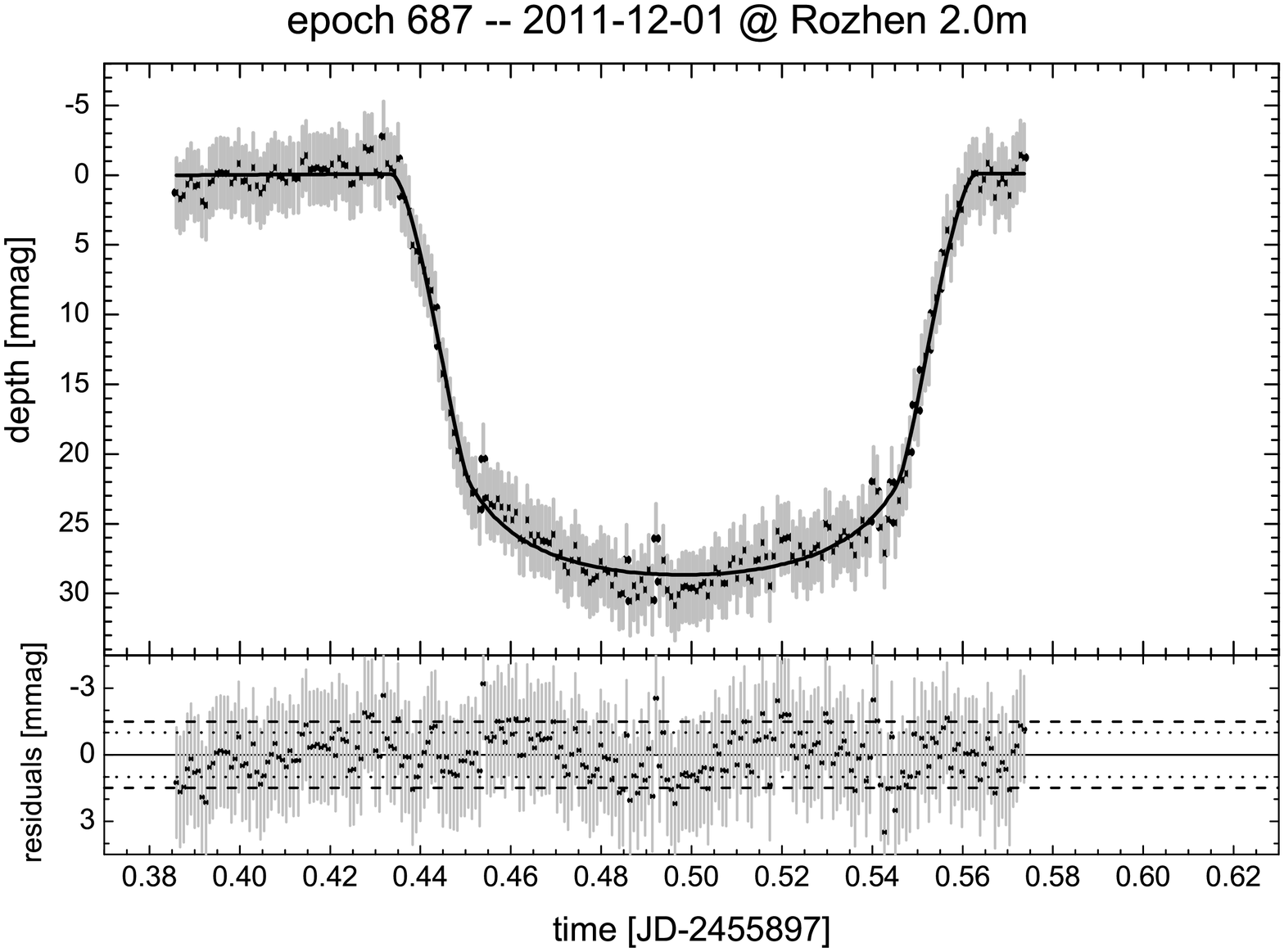}
  \includegraphics[width=0.29\textwidth]{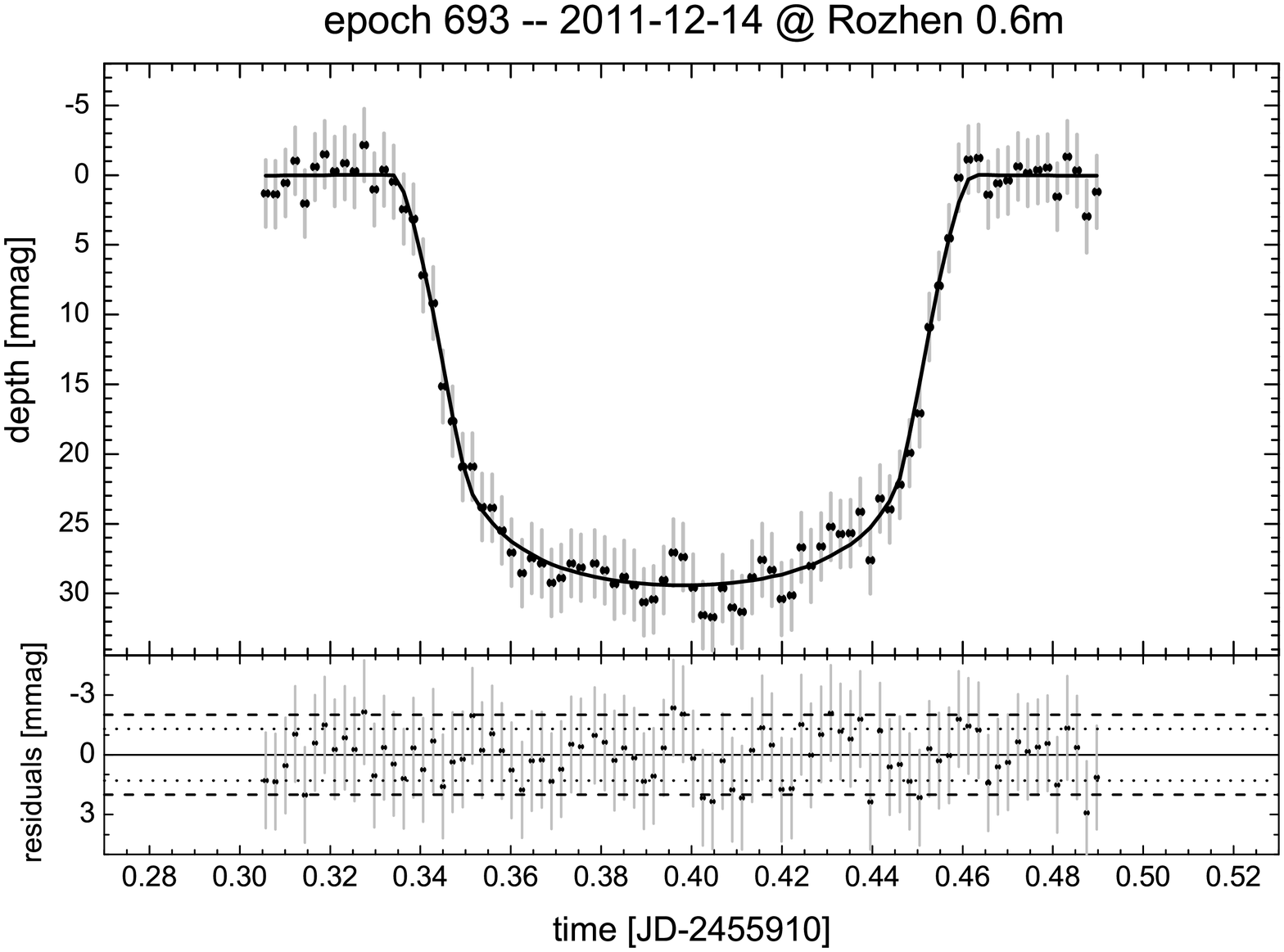}
  \includegraphics[width=0.29\textwidth]{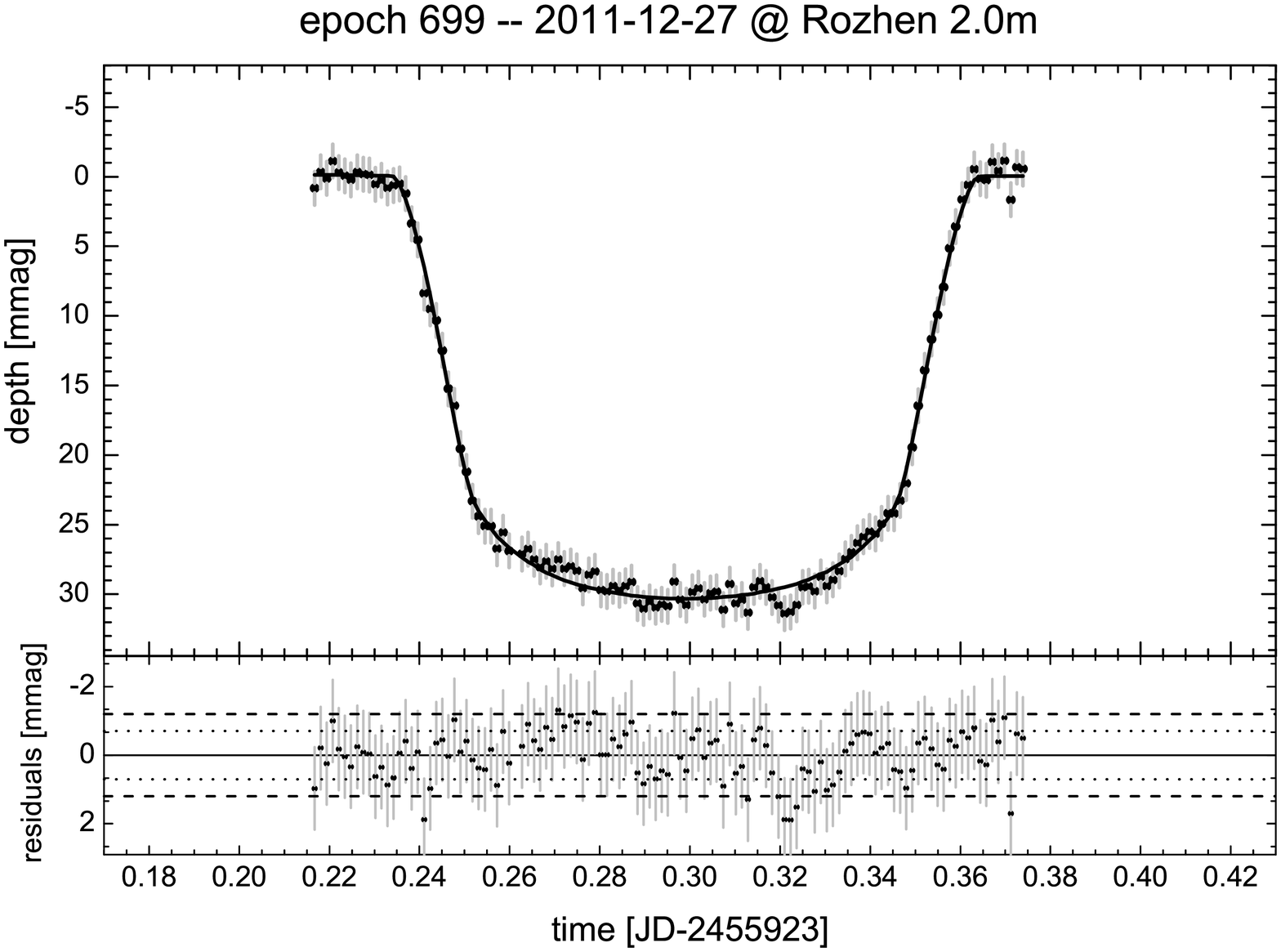}

  \includegraphics[width=0.30\textwidth]{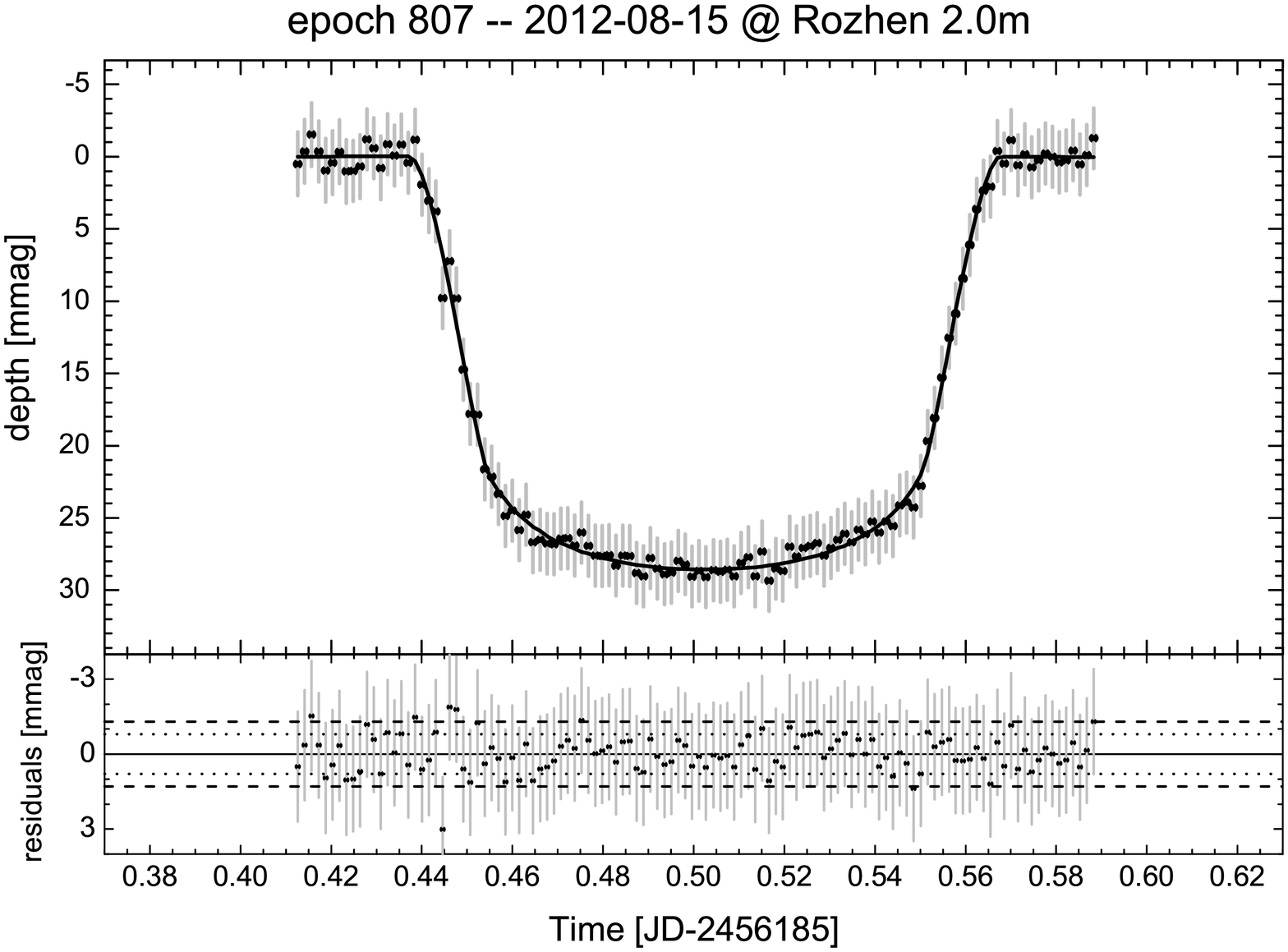}
  \includegraphics[width=0.30\textwidth]{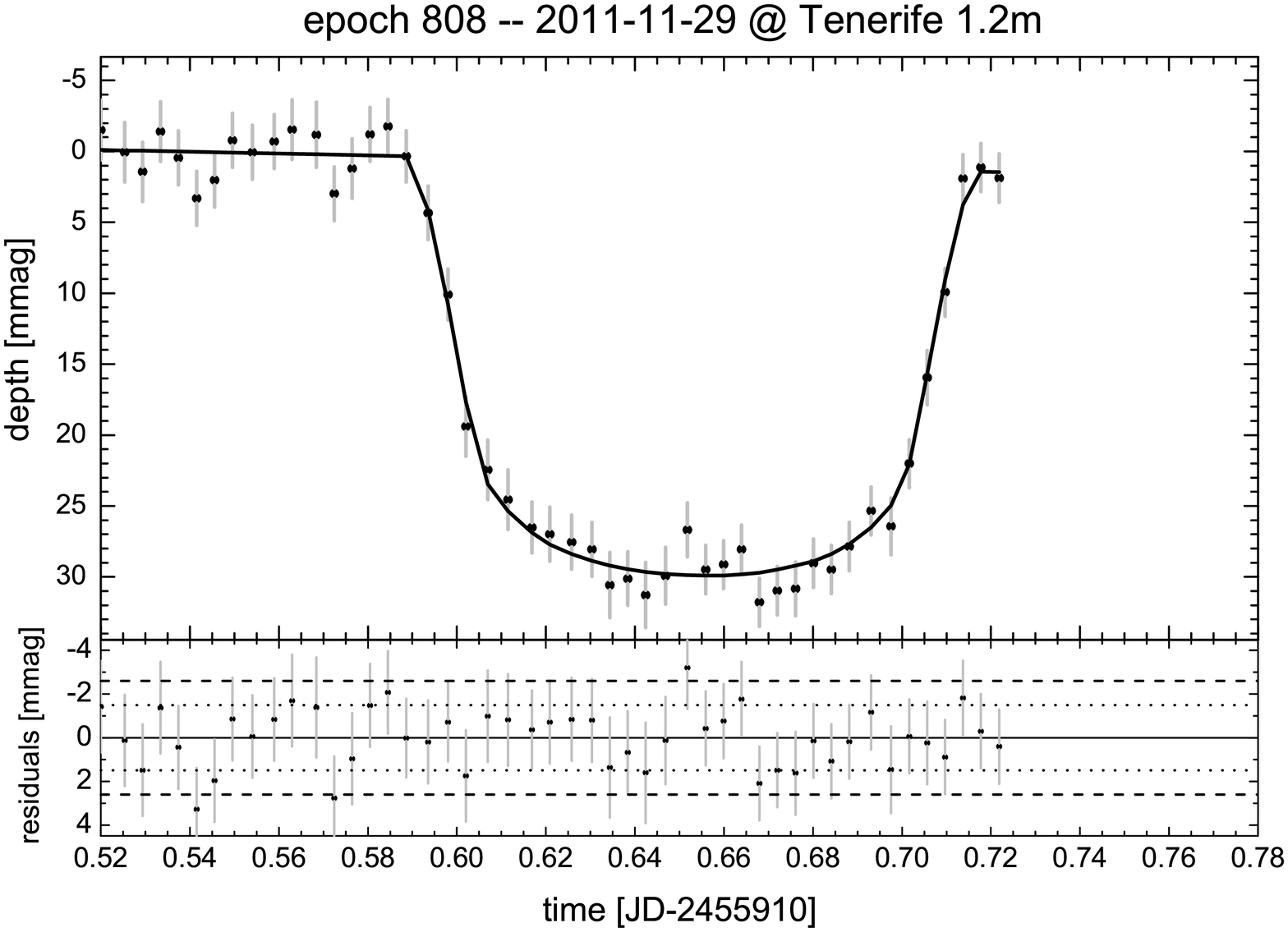}
  \includegraphics[width=0.30\textwidth]{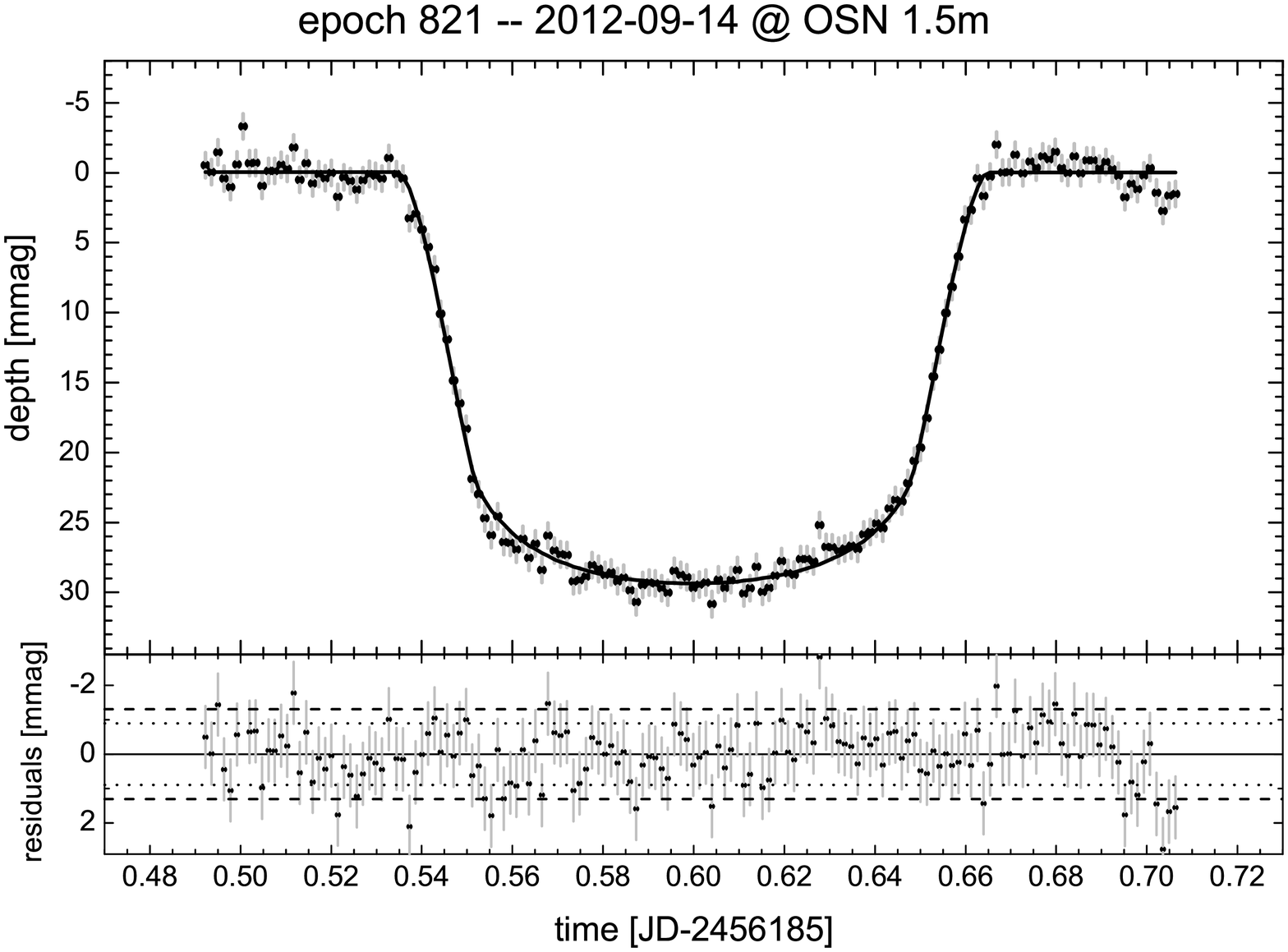}

  \includegraphics[width=0.30\textwidth]{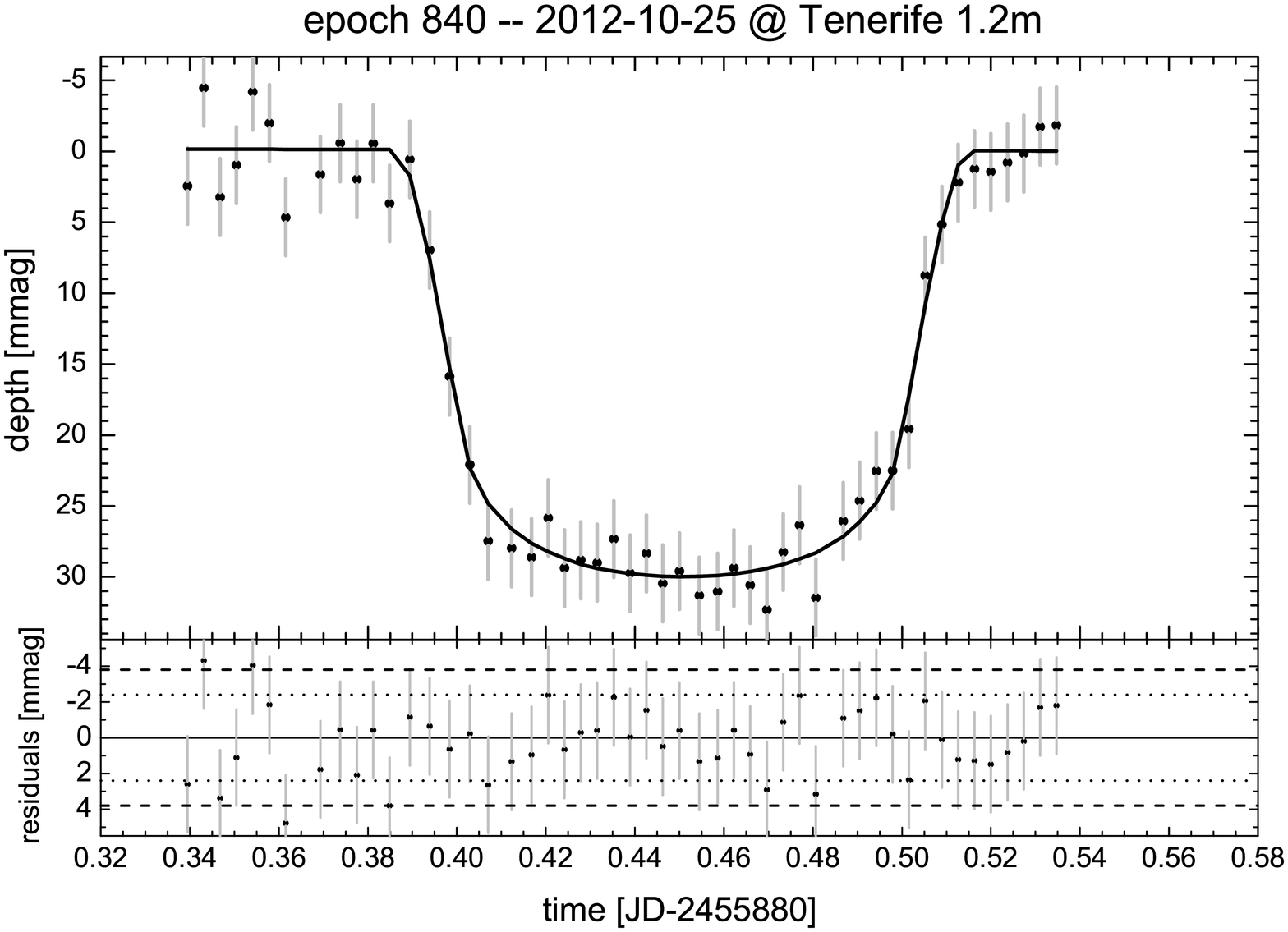}
  \includegraphics[width=0.30\textwidth]{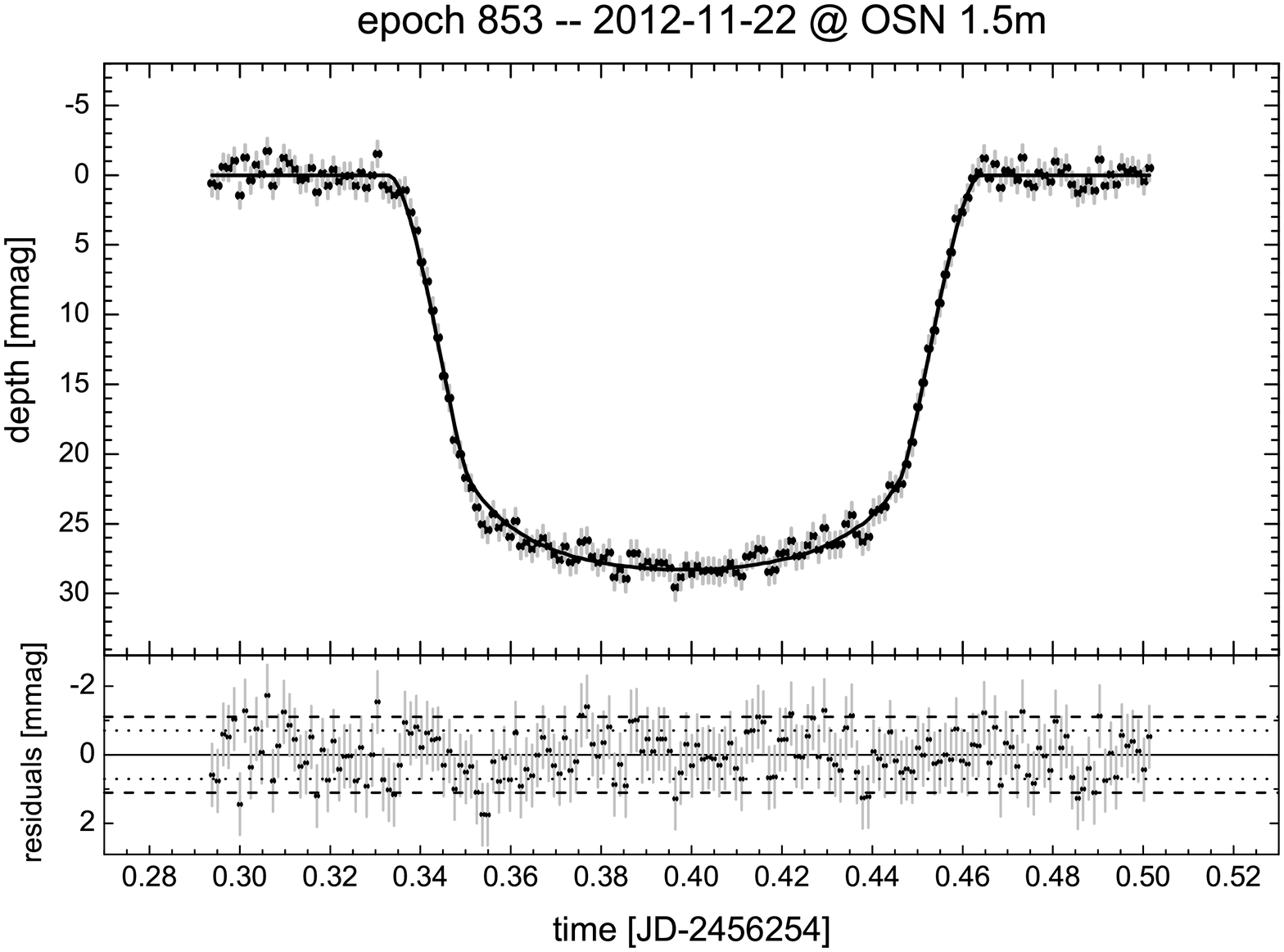}
  \includegraphics[width=0.30\textwidth]{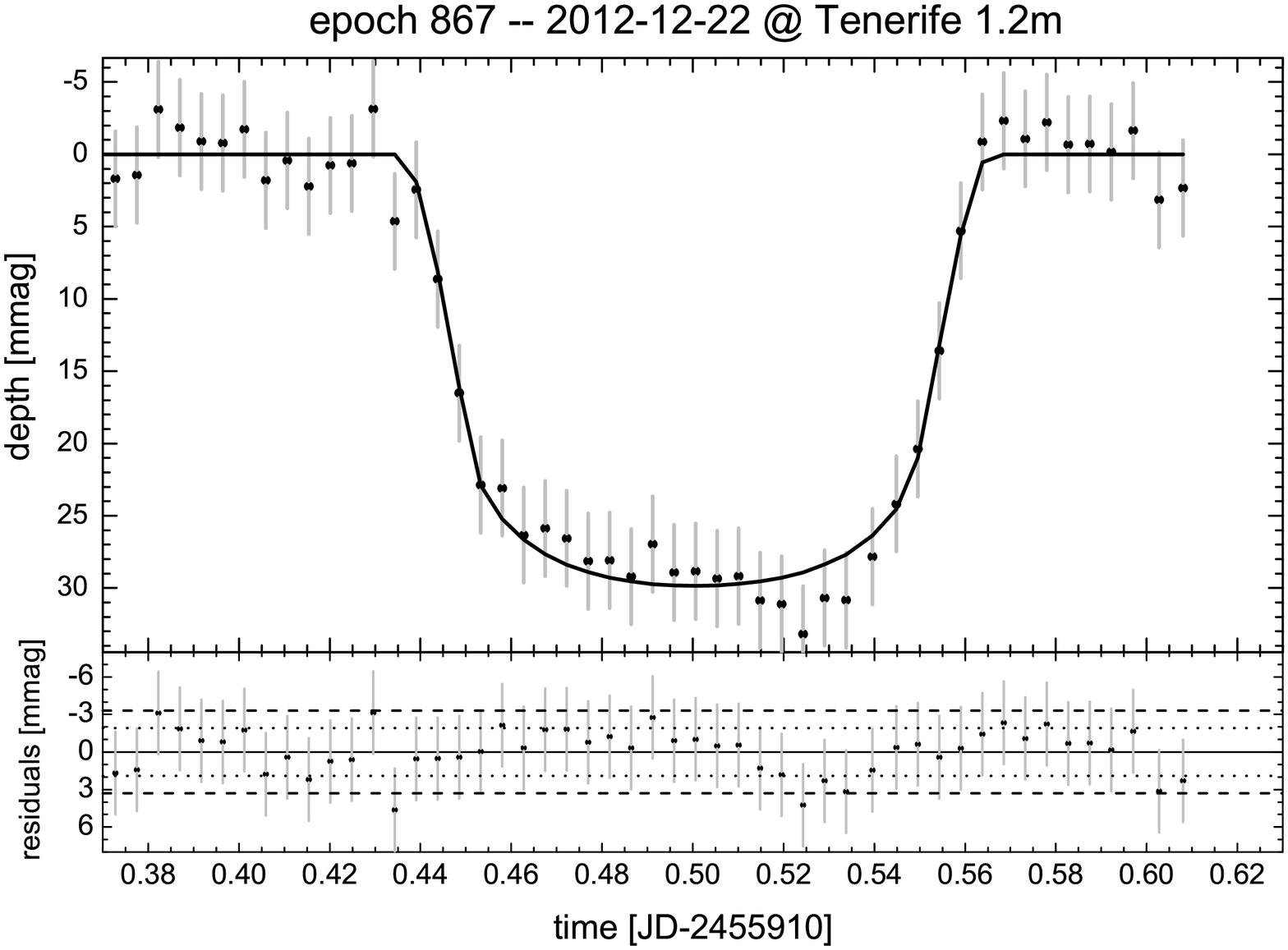}

  \includegraphics[width=0.30\textwidth]{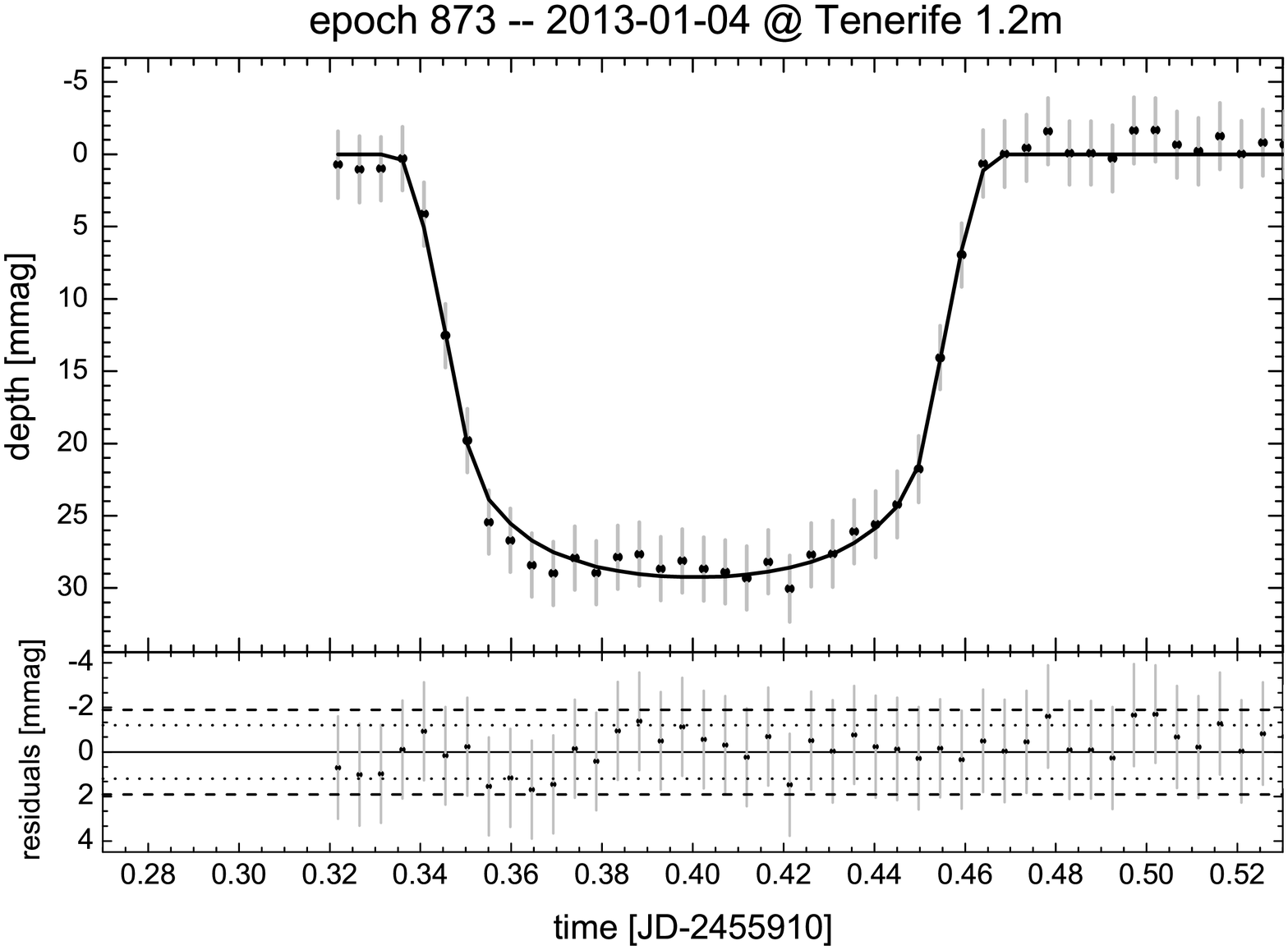}
  \includegraphics[width=0.30\textwidth]{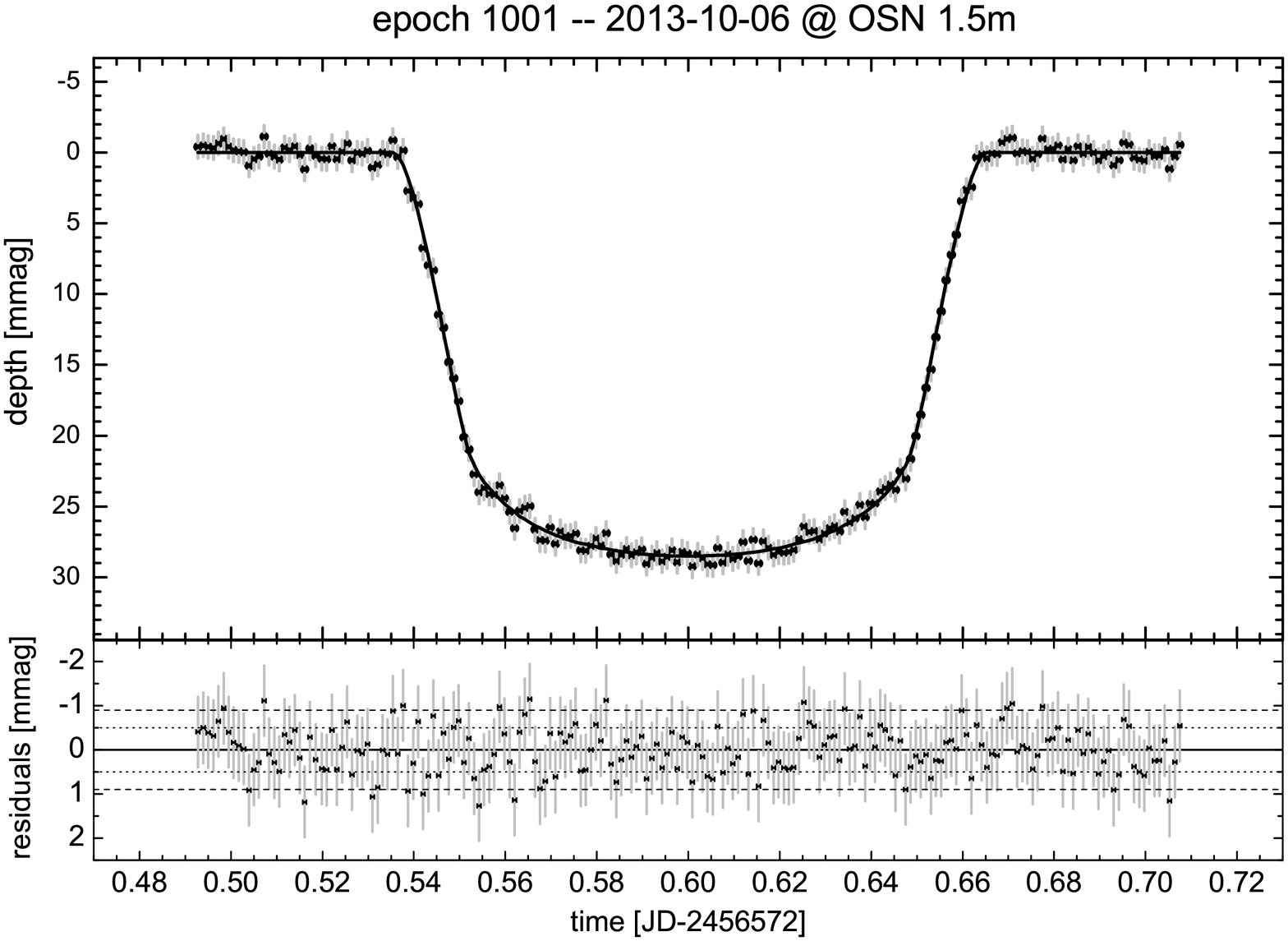}
  \includegraphics[width=0.30\textwidth]{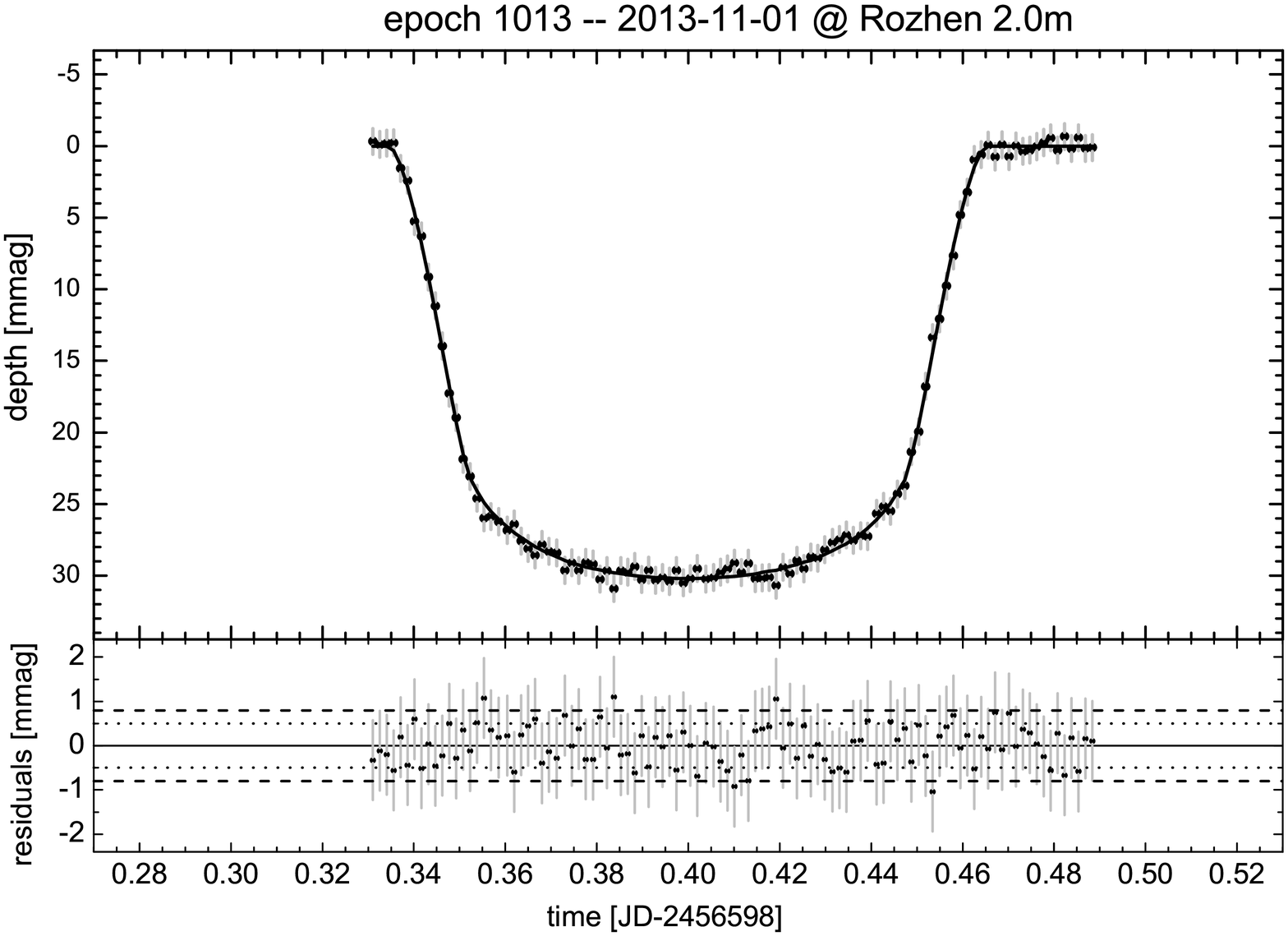}
  
  \includegraphics[width=0.30\textwidth]{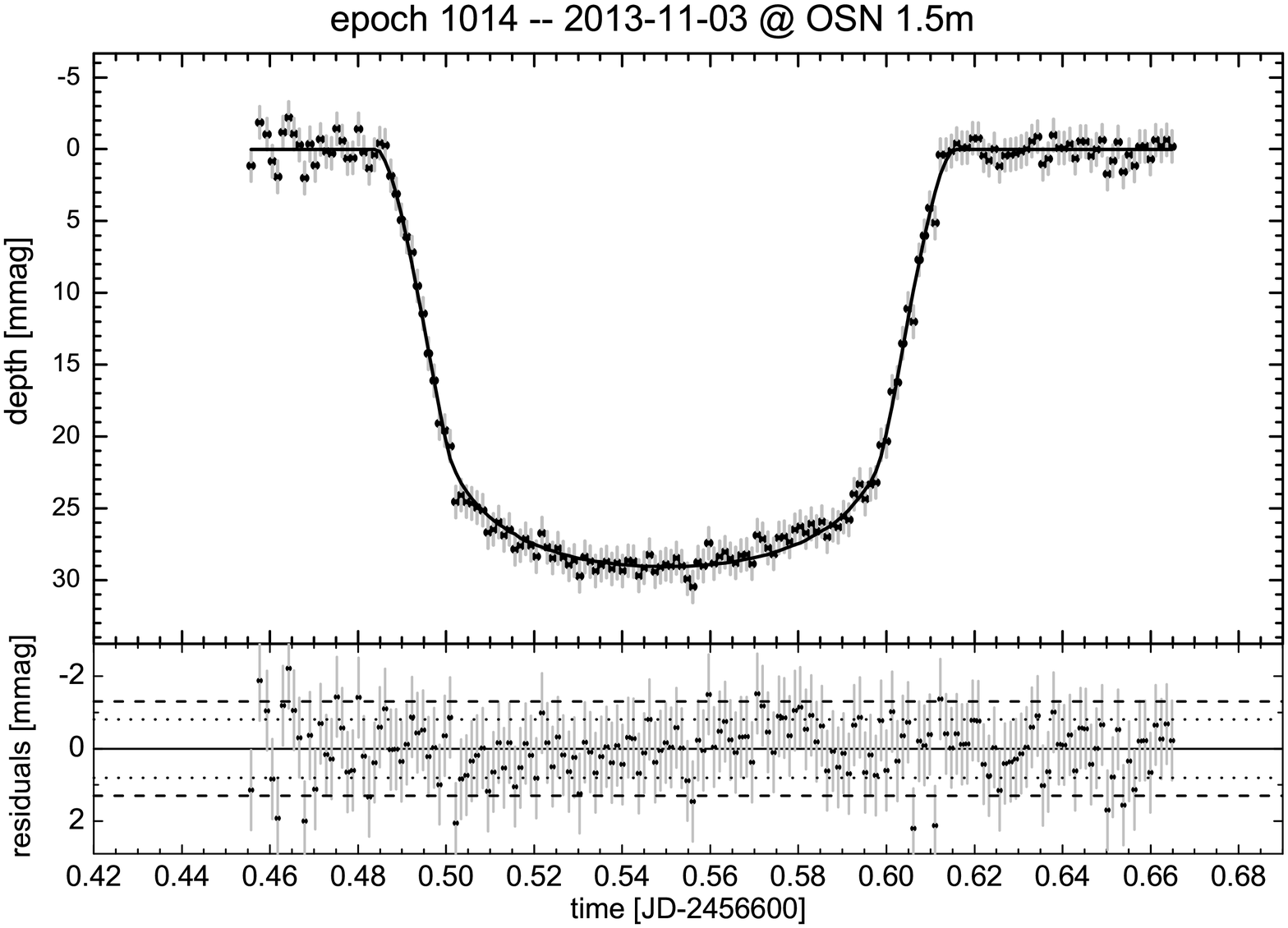}
  \includegraphics[width=0.30\textwidth]{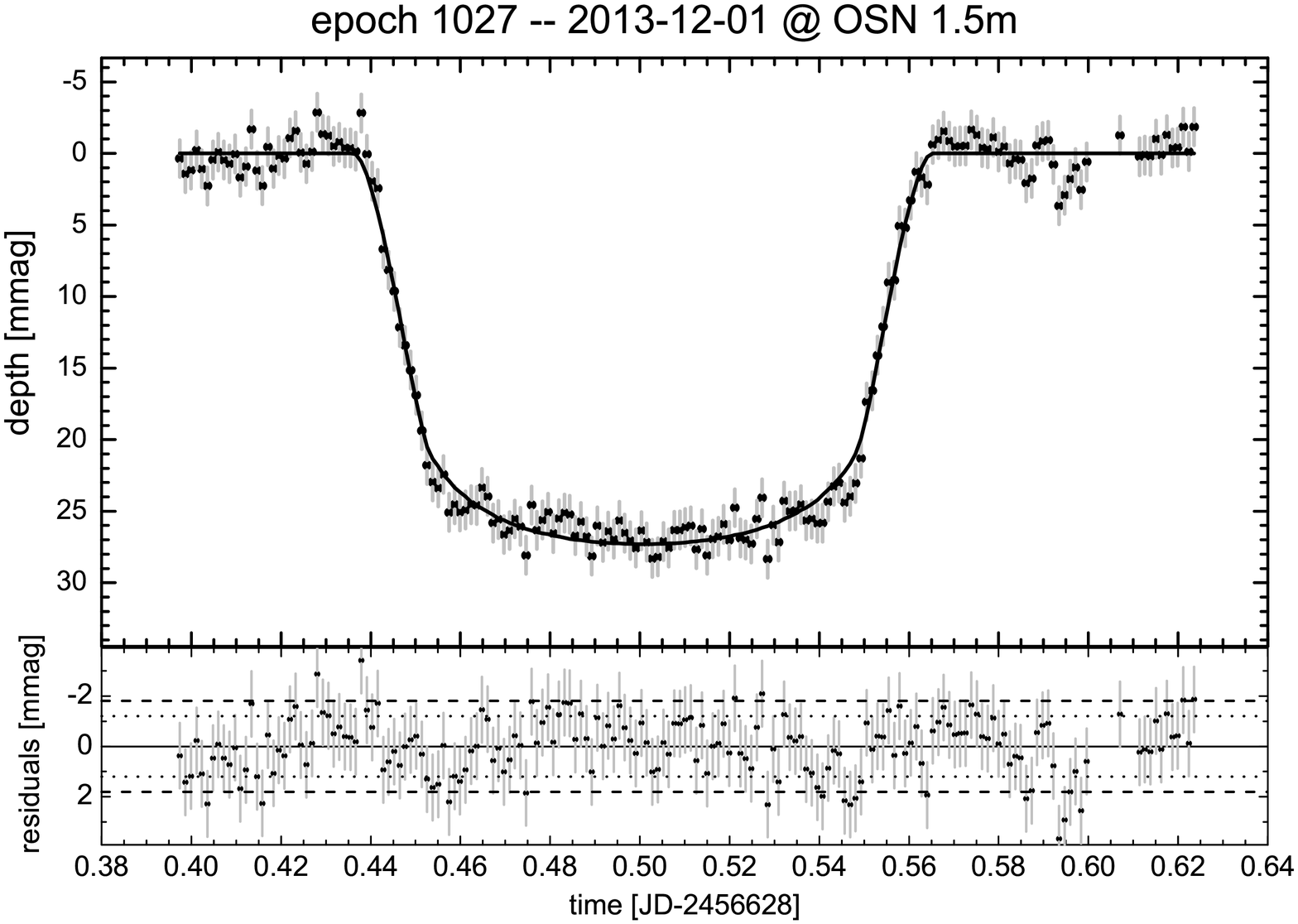}
  \includegraphics[width=0.30\textwidth]{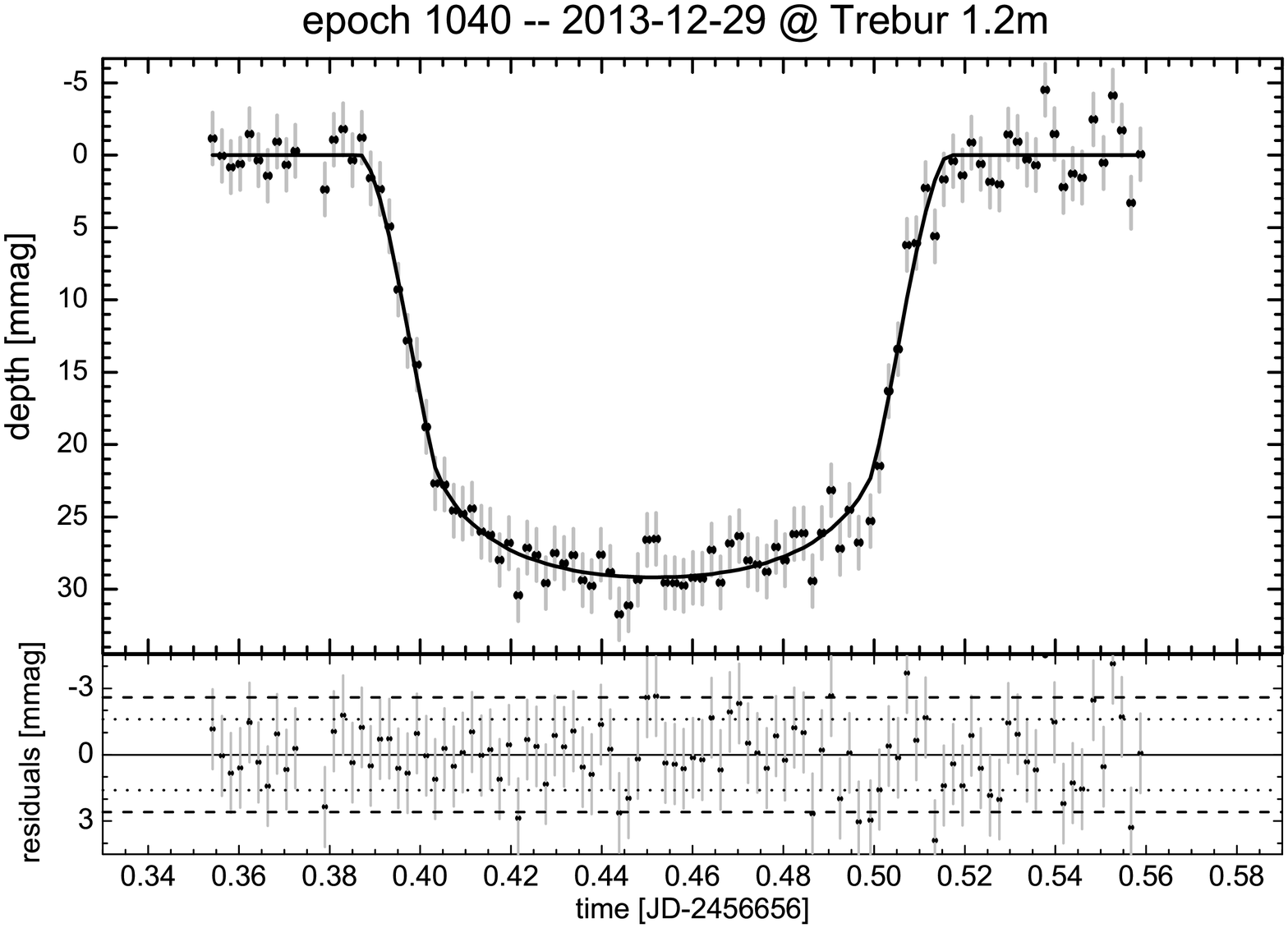}

  \caption{The threefold binned transit light curves of those 18 complete transit event observations, where no simultaneous observations 
at different telescopes could be achieved. The upper panels show the light curve, the lower panels show the residuals. The rms of the 
fit in the original (dashed lines) and the threefold binned light curve (dotted lines) are shown as well.}\label{fig:TransitLightCurves1} 
\end{figure*}

\begin{figure*}
  \includegraphics[width=0.42\textwidth]{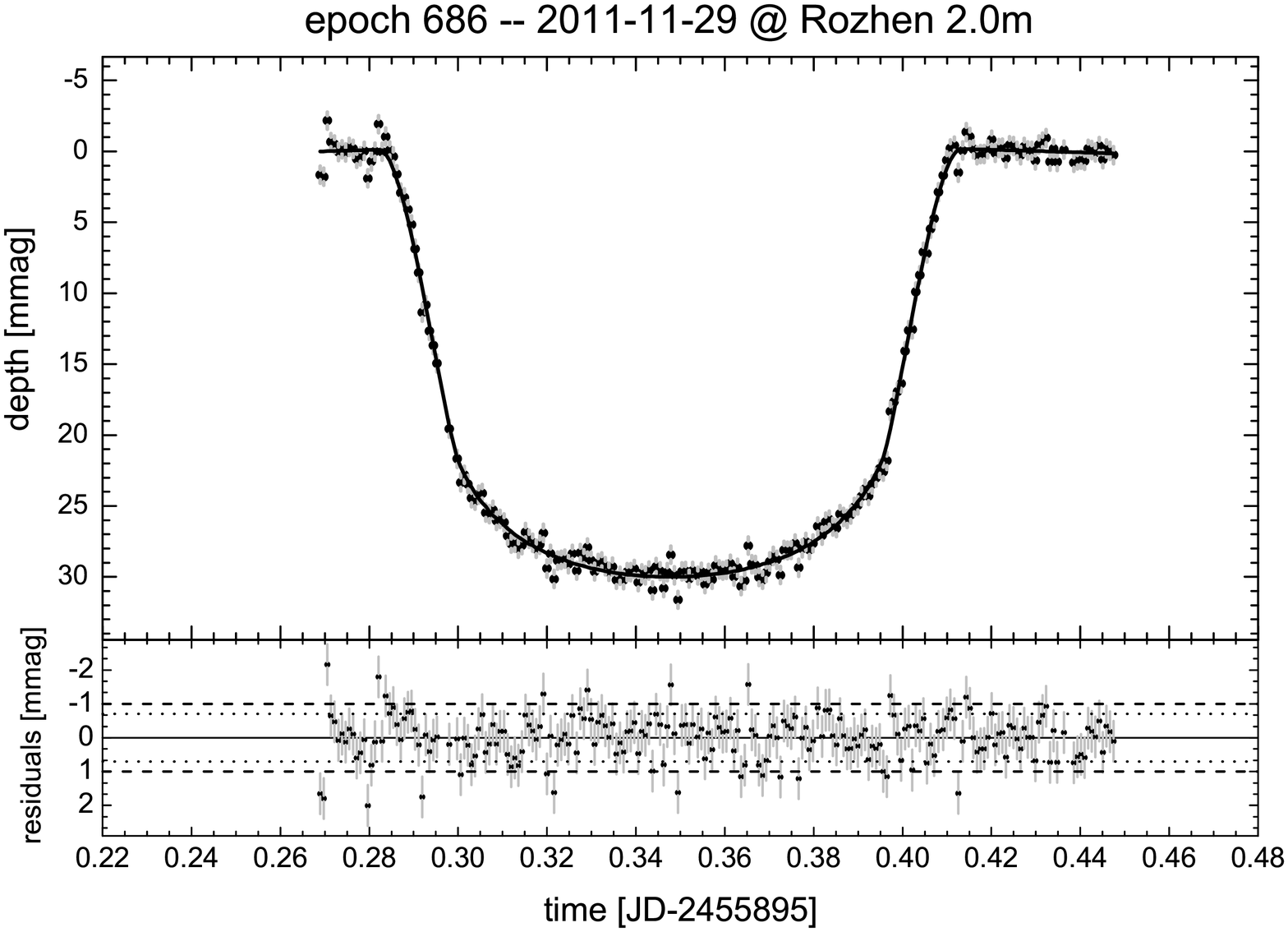}
  \includegraphics[width=0.42\textwidth]{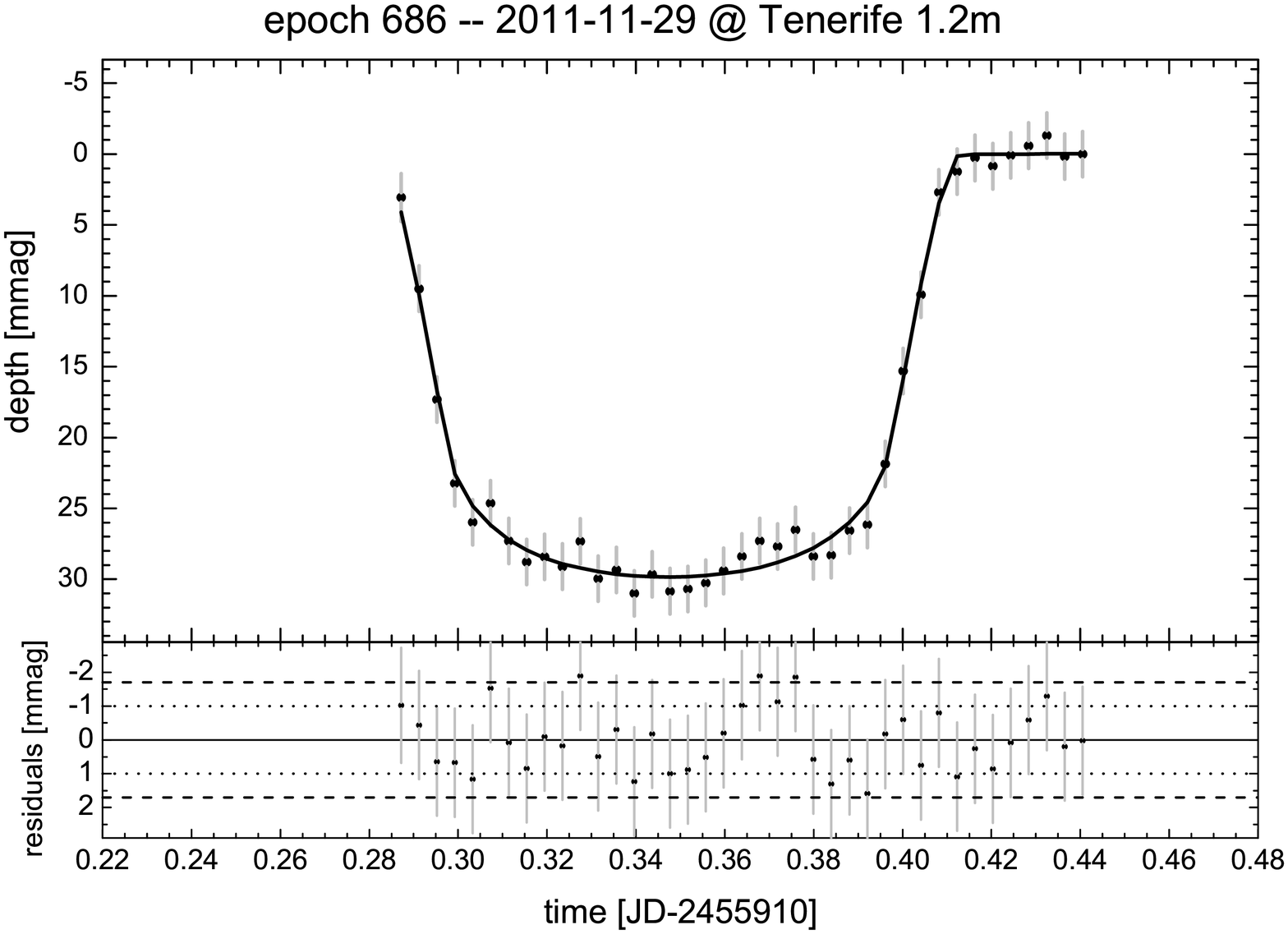}

  \includegraphics[width=0.42\textwidth]{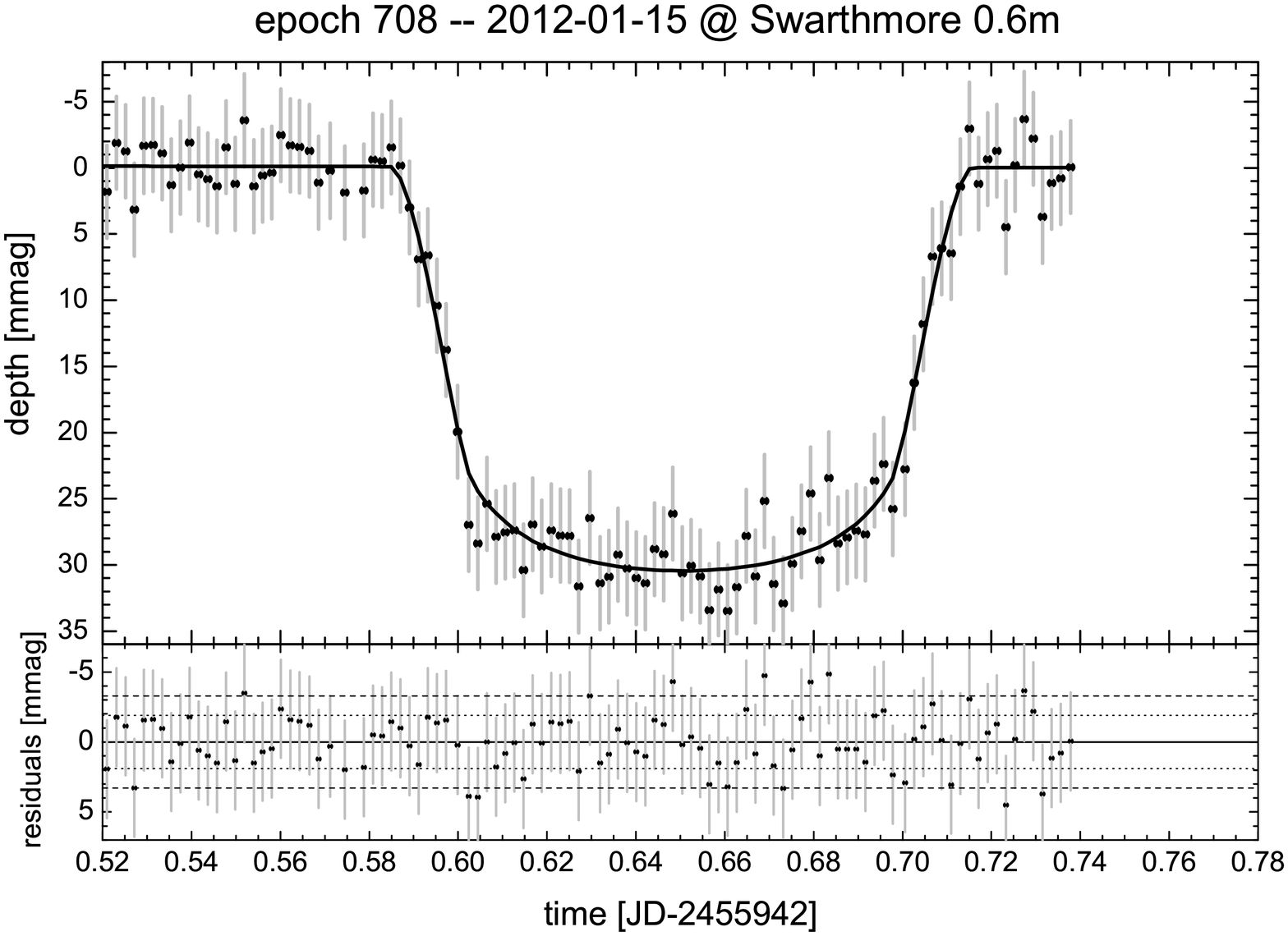}
  \includegraphics[width=0.42\textwidth]{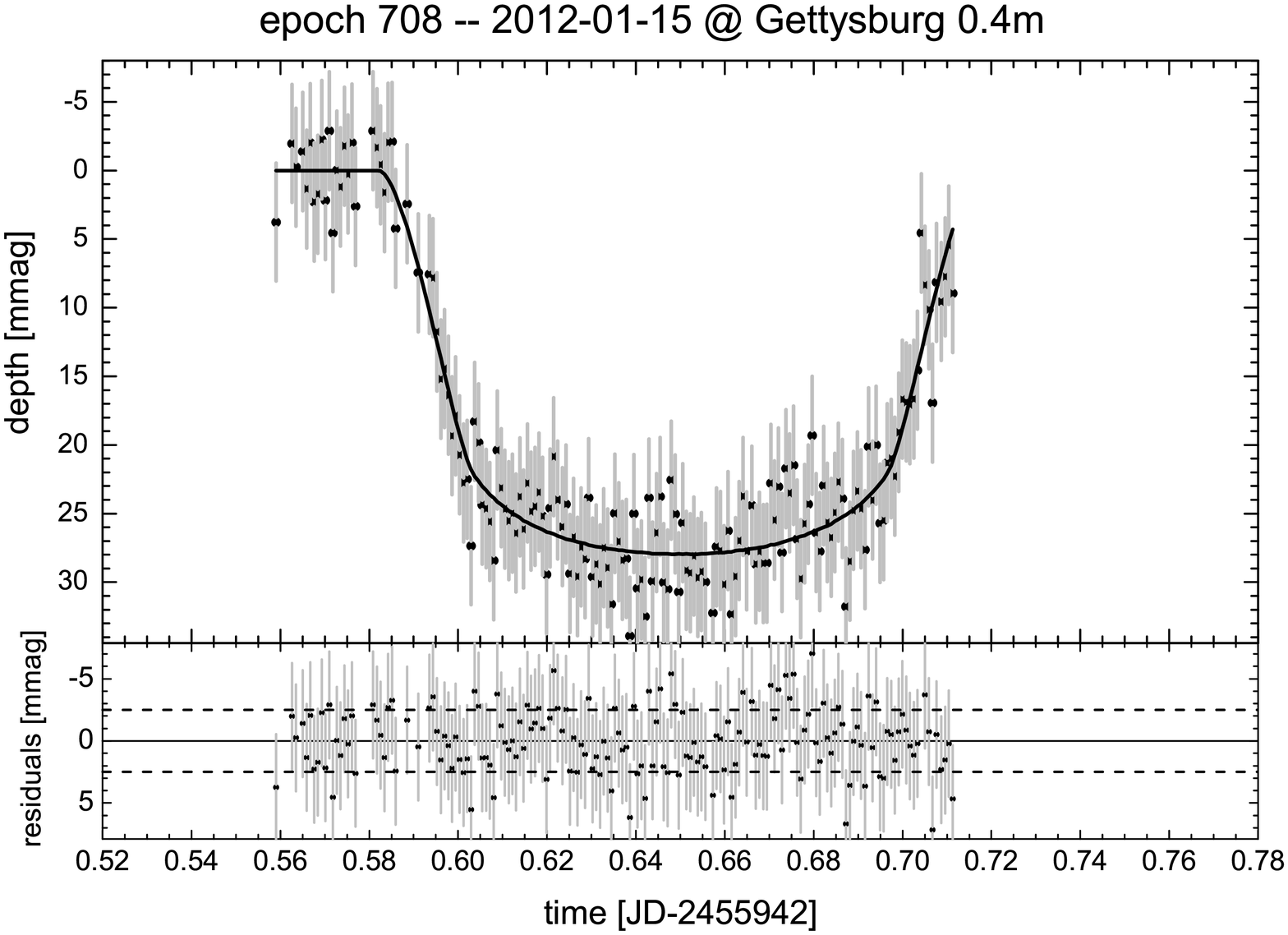}

  \includegraphics[width=0.42\textwidth]{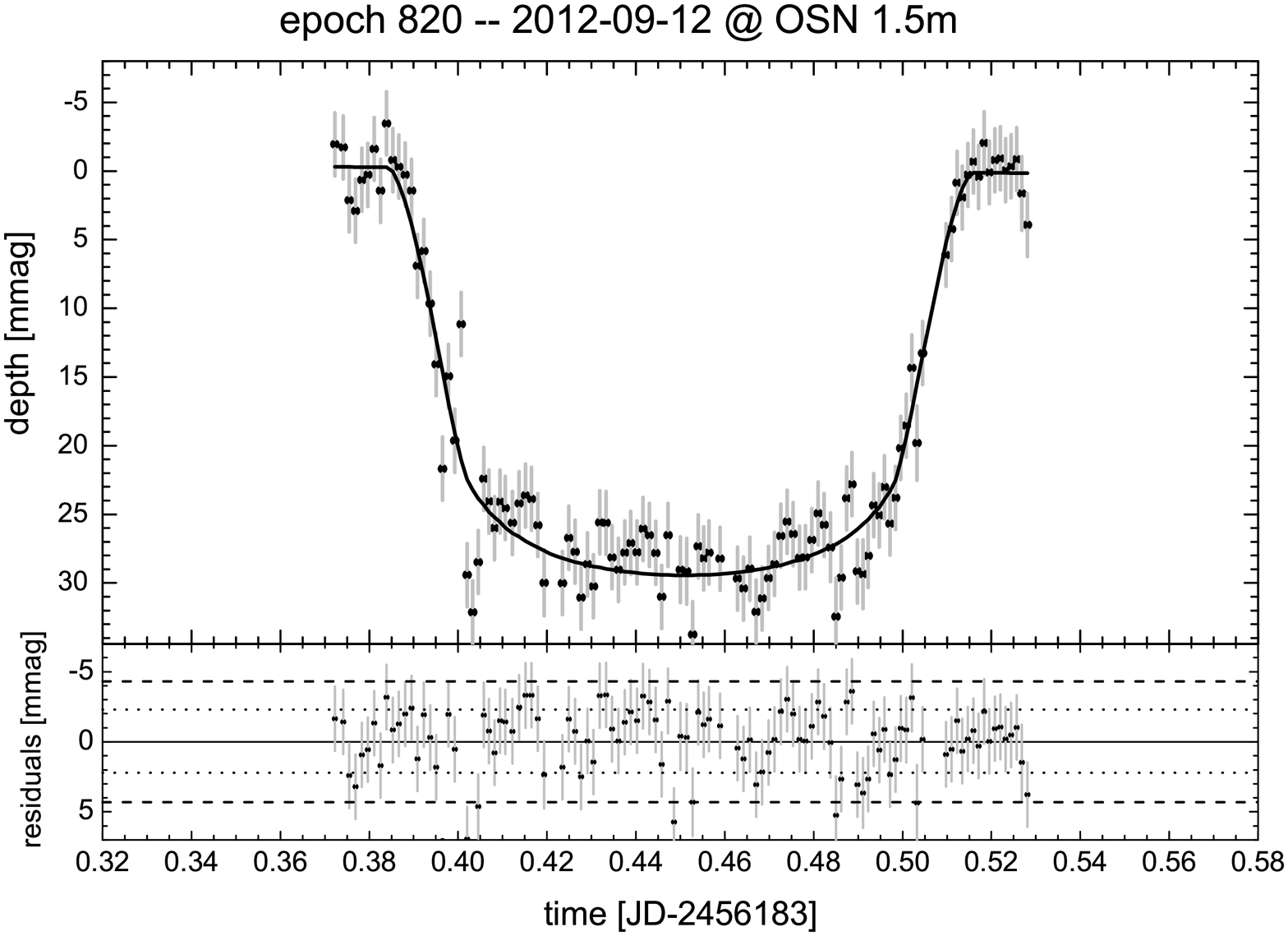}
  \includegraphics[width=0.42\textwidth]{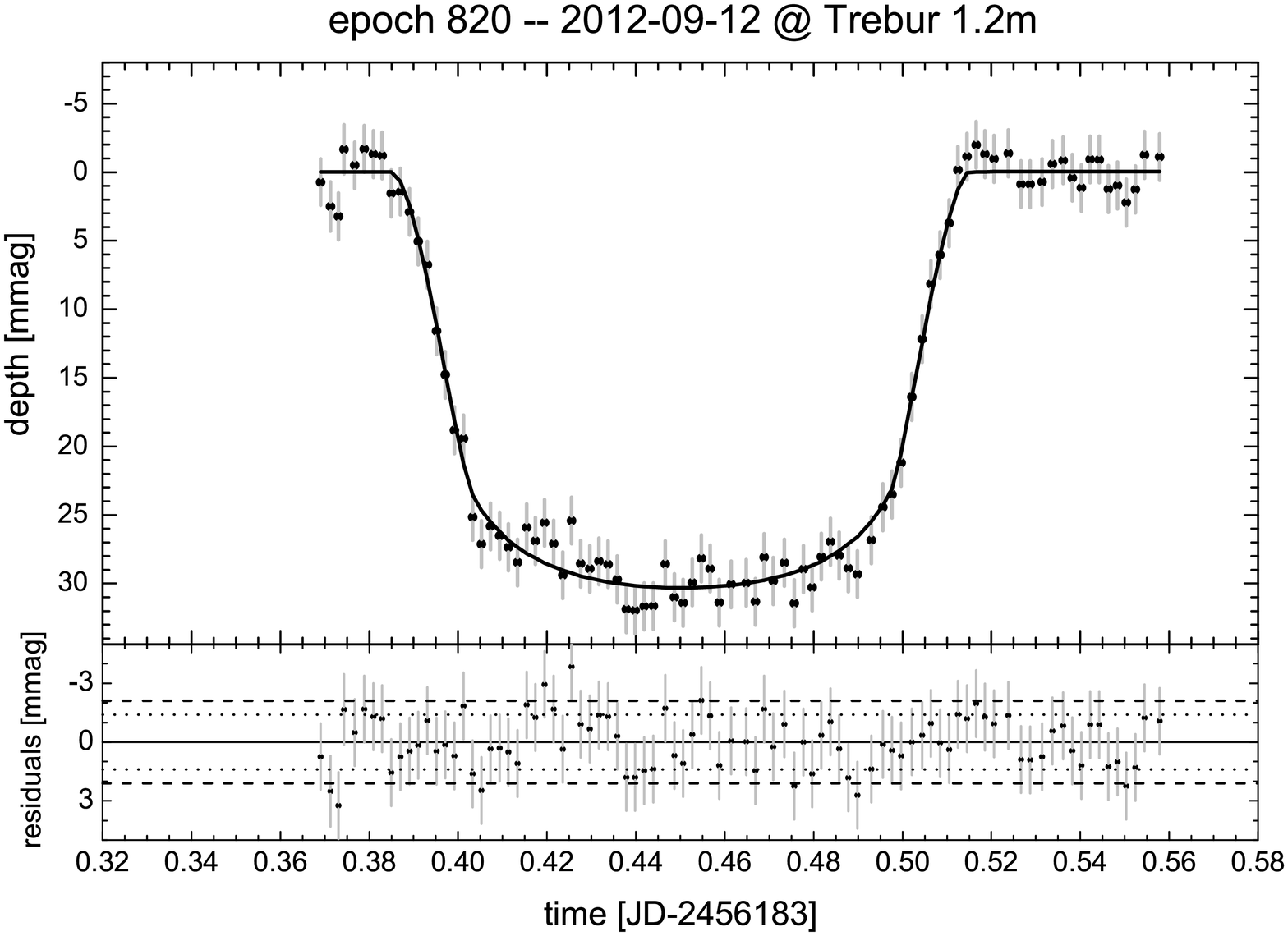}

  \includegraphics[width=0.42\textwidth]{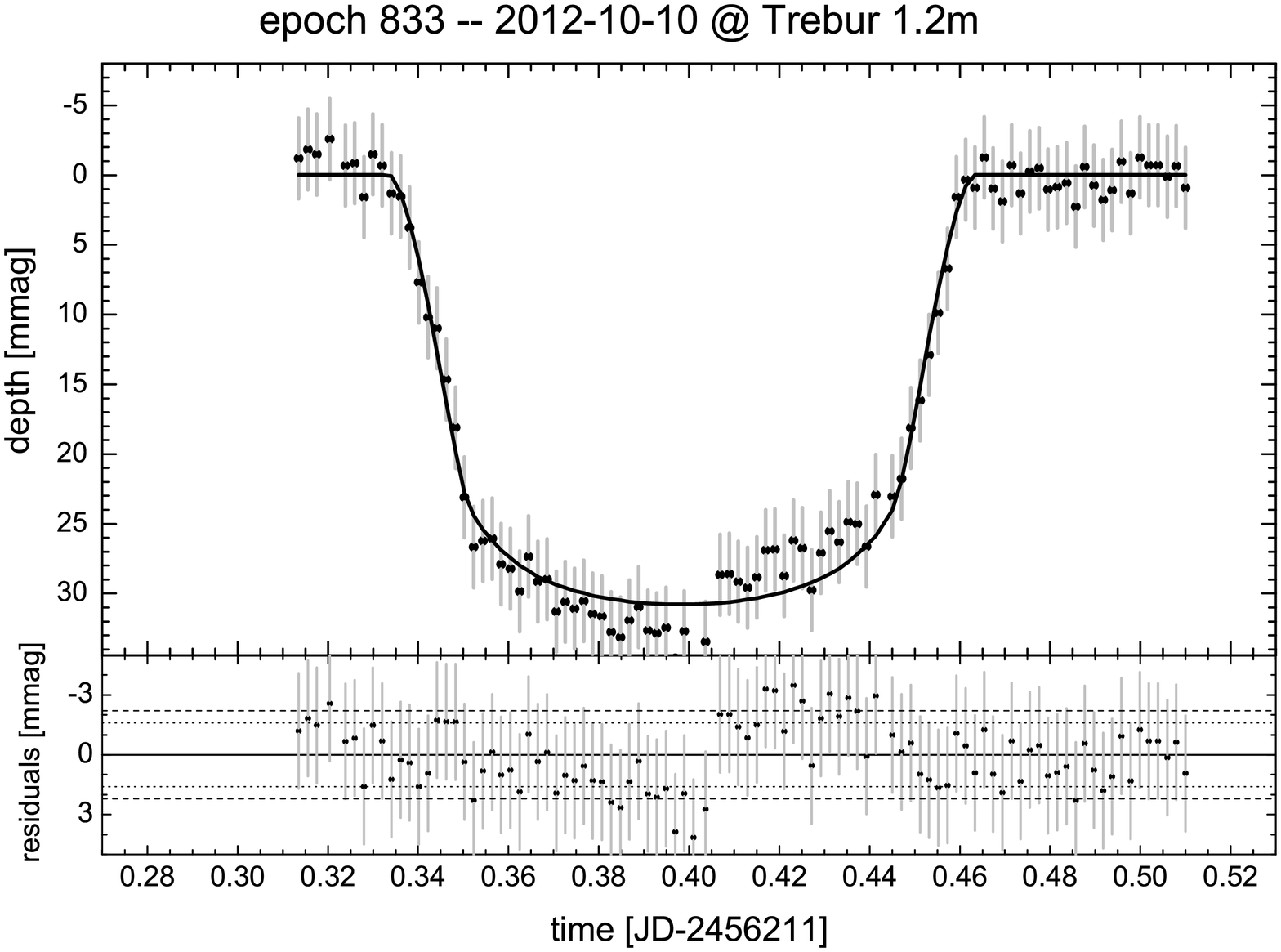}
  \includegraphics[width=0.42\textwidth]{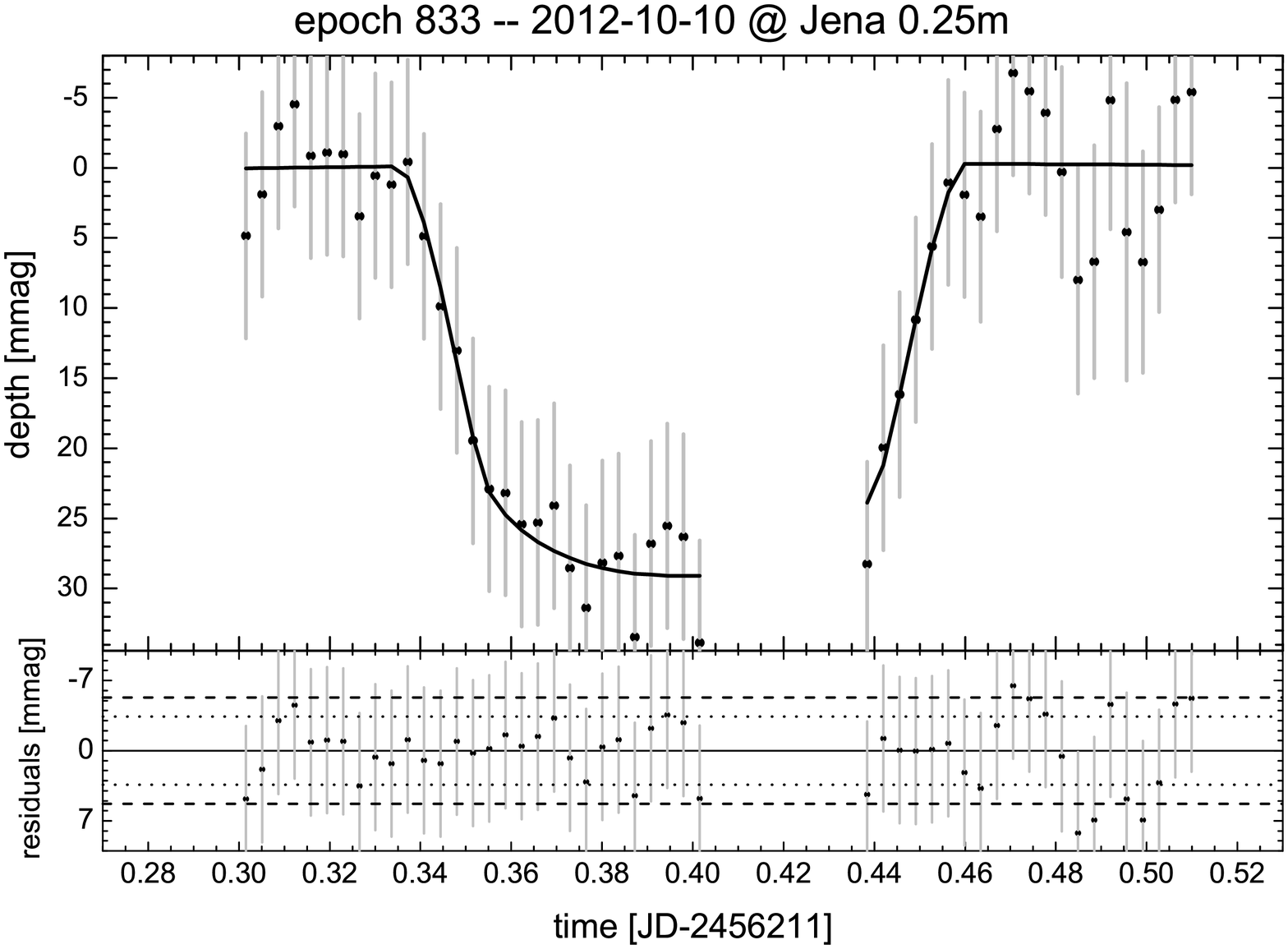}

  \caption{The threefold binned transit light curves of those four transit event observations, where simultaneous observations 
at two different telescopes could be achieved. The upper panels show the light curve, the lower panels show the residuals. The 
rms of the fit in the original (dashed lines) and the binned light curve (dotted lines) are shown as well.}\label{fig:TransitLightCurves2}
\end{figure*}

\begin{figure*}
  \includegraphics[width=0.32\textwidth]{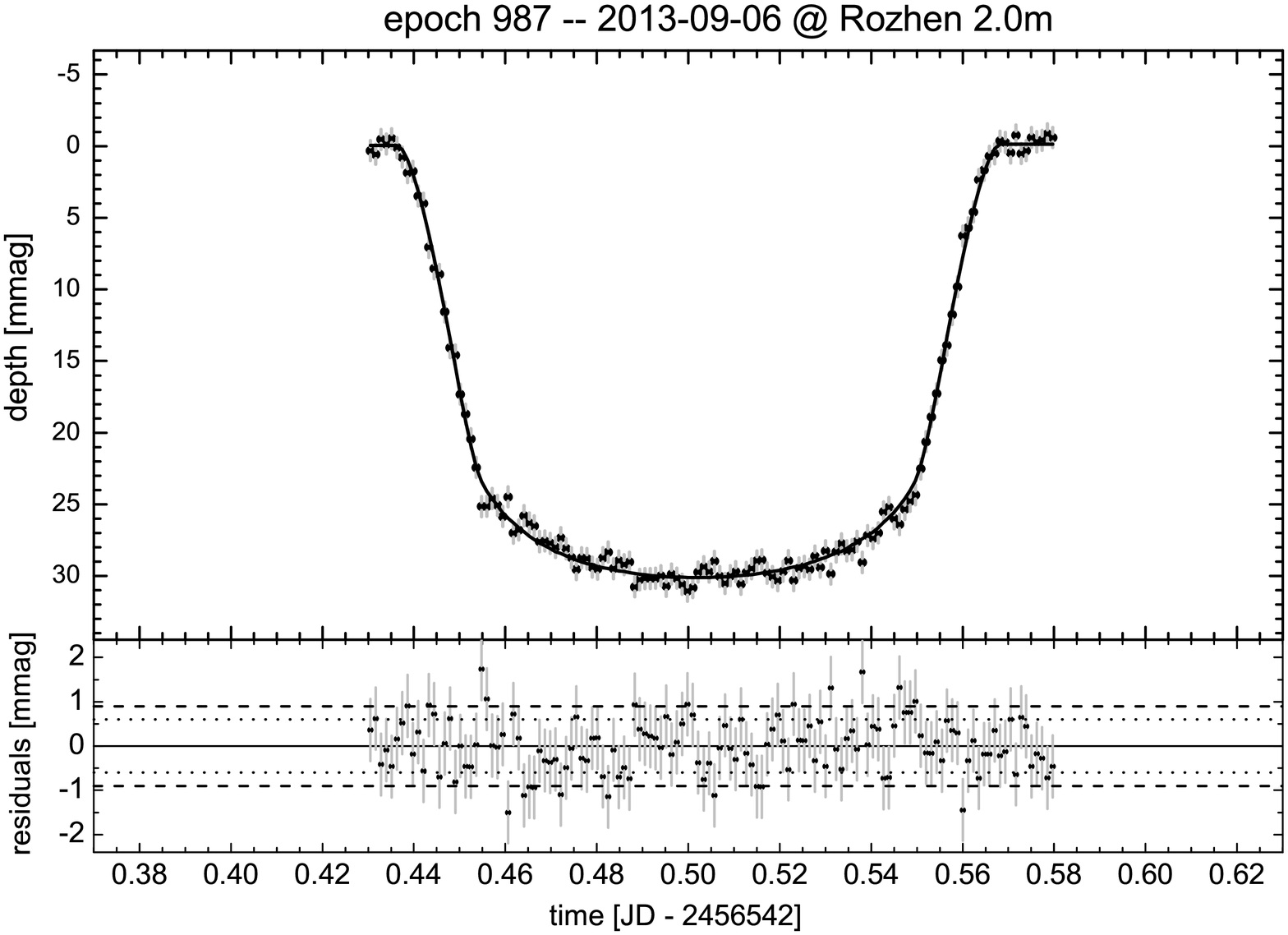}
  \includegraphics[width=0.32\textwidth]{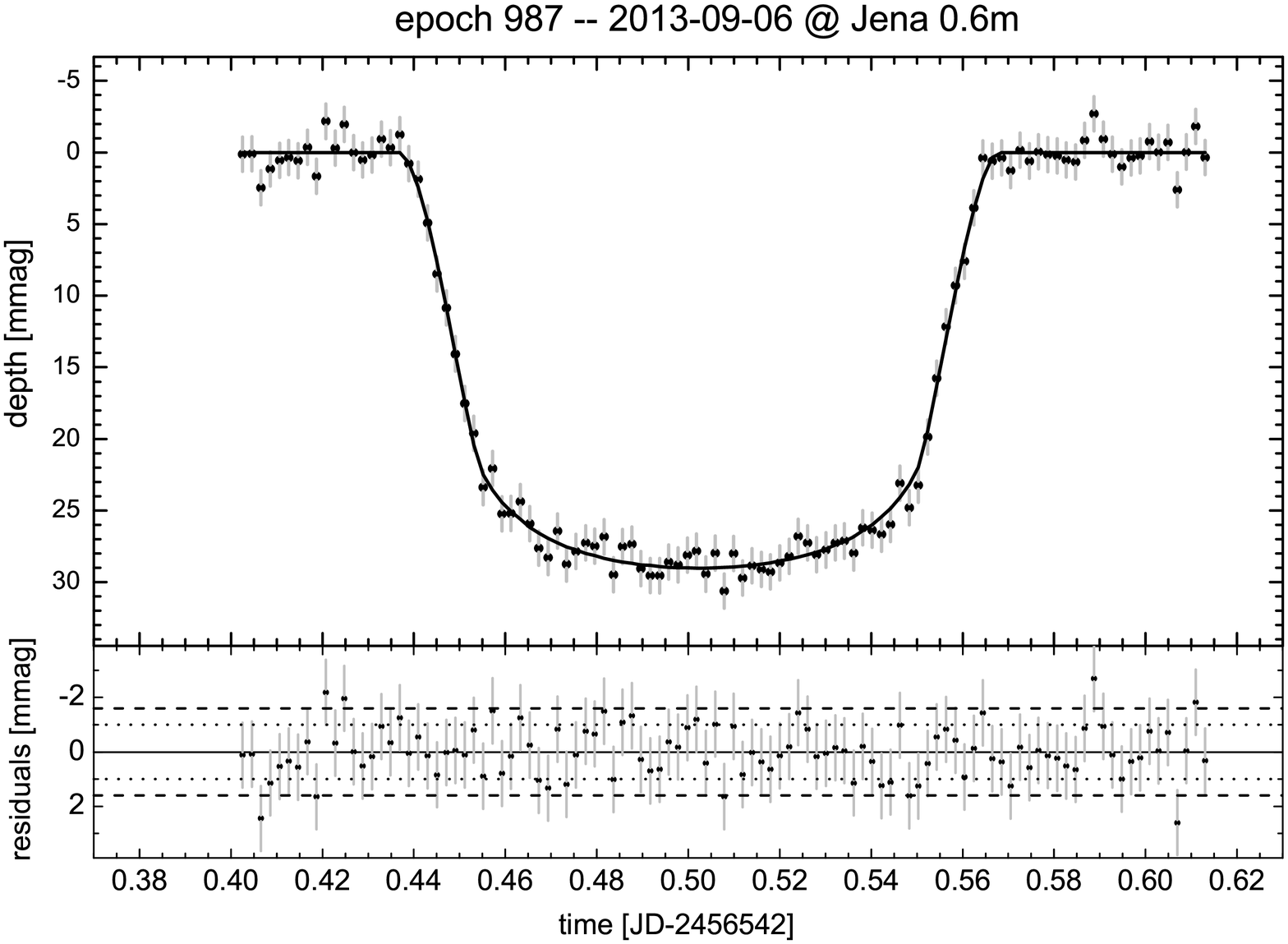}
  \includegraphics[width=0.32\textwidth]{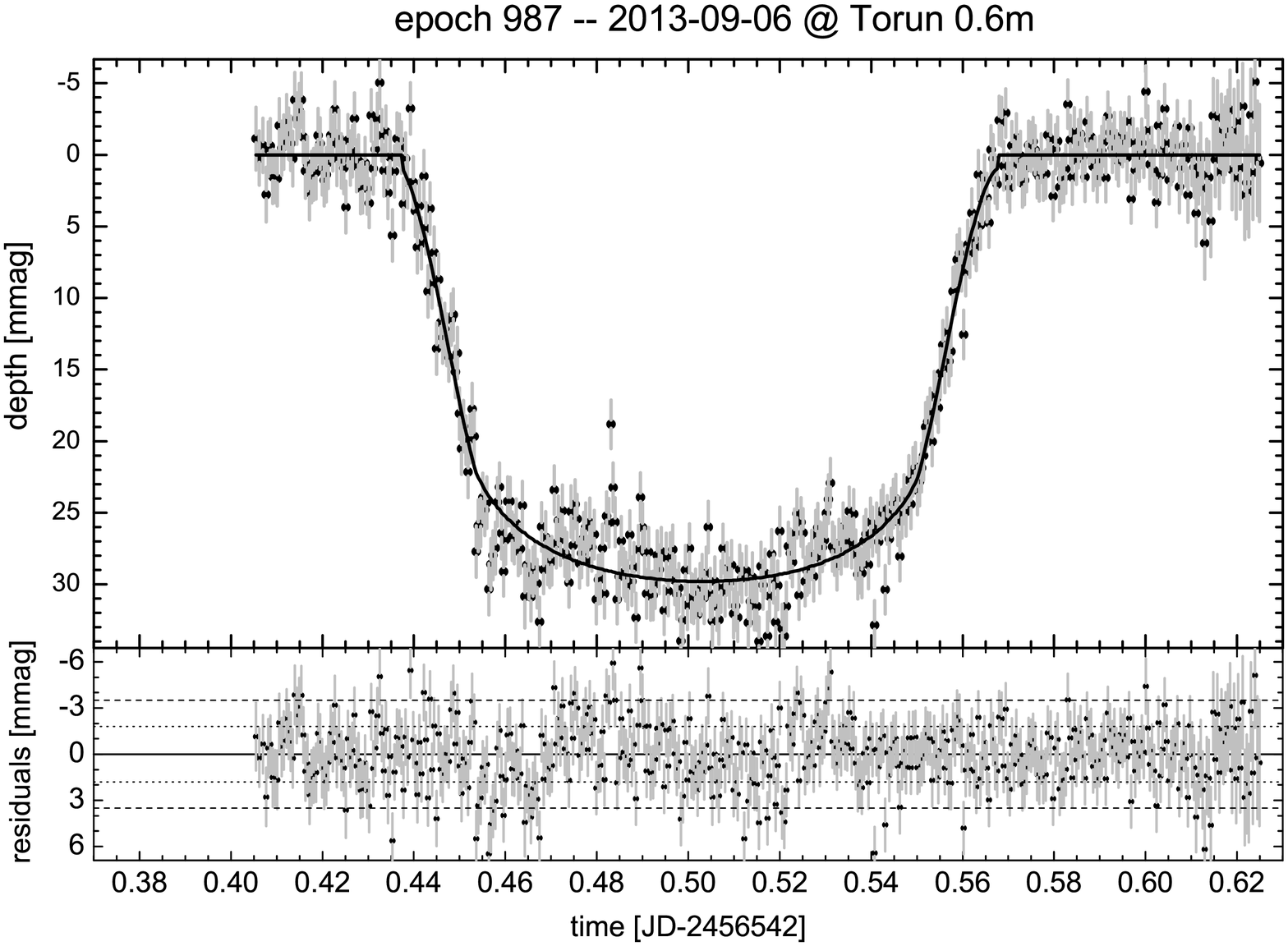}
  \caption{The threefold binned transit light curves of the transit event on 2013 September 6. The transit 
was observed from three different observatories simultaneously: Rozhen 2.0m, Jena 0.6m, and Torun 0.6m. The upper panels 
show the light curve, the lower panels show the residuals. The rms of the fit in the original (dashed lines) and the binned 
light curve (dotted lines) are shown as well.}\label{fig:TransitLightCurves3}
\end{figure*}

\begin{figure}
   \includegraphics[width=1\columnwidth]{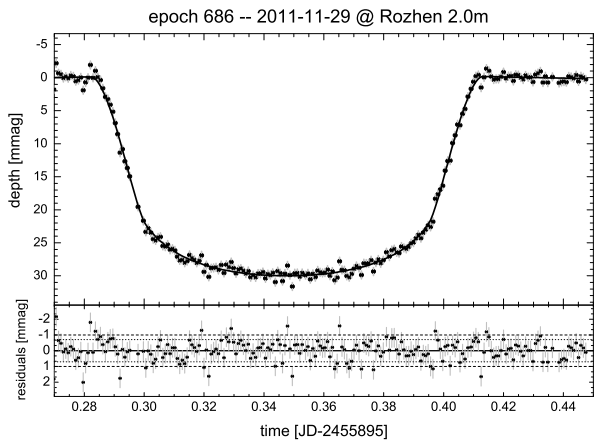}
  \caption{ The best transit light curve obtained for this program so far observed with the 2m Rozhen telescope. The rms of the fit is 
1.0 and 0.7 mmag in the unbinned and binned light curve, respectively. The mid-transit time has a fitting precision of 14~s.}\label{fig:BestTransitLightCurves}
\end{figure}

After fitting an individual transit with both \textit{JKTEBOP} and \textit{TAP} there are (typically) eight different fit results 
(two programs with fixed and free limb-darkening using binned and unbinned data, respectively) with in total 28 individually 
derived error bars for the properties $i$, $R_\textrm{p}/R_\textrm{s}$, $T_{\textrm{mid}}$, and $r_\textrm{p}+r_\textrm{s}$ or $a/R_\textrm{s}$.

Regarding all properties, the fit results for the original light curves and the binned ones show no 
significant differences. Especially concerning the precision of the mid-transit time the results can not be improved by binning 
the light curve. Though one can reduce the error of an individual data point, one reduces the timing resolution and hence decreases 
the timing precision \citep[as well as the overall fitting precision; discussed in detail in e.g.][]{Kipping2010}. 
Thus, to get one final result for each transit event, we first convert $r_\textrm{p}+r_\textrm{s}$ to $a/R_\textrm{s}$ and then 
take the average of all obtained values.

In all cases the spread of the fitted values is smaller than the averaged error bar, hence the differences between the models are 
smaller than the fitting precision. This result is in good agreement with those of e.g. \citet*{Hoyer2012}. Nevertheless, as 
previously discussed in e.g. \citet{Maciejewski2013a} and \citet{Carter2009}, \textit{JKTEBOP} may 
underestimate the error bars. Though, in this work the differences between the errors derived by \textit{JKTEBOP} (Monte Carlo, 
residual-shift, and bootstrapping) and those of \textit{TAP} (MCMC) are not as noticeable as in e.g. \citet{Maciejewski2013a} 
(especially for $i$, $R_\textrm{p}/R_\textrm{s}$, and $a/R_\textrm{s}$, but larger for $T_{\textrm{mid}}$), the mean errors derived 
by the \textit{TAP}-code are used as final errors.

Keeping the LD coefficients fixed at their theoretical values does not result in significant differences of the light curve fit. This is 
true for the value as well as the error estimations.

The final light curves and the model fits are shown in Fig.~\ref{fig:TransitLightCurves1} for the single site, and 
Fig.~\ref{fig:TransitLightCurves2}, and \ref{fig:TransitLightCurves3} for the multi site observations. Our most precise light curve has been obtained
using the Rozhen 2.0m telescope and is shown in Fig.~\ref{fig:BestTransitLightCurves}.

\subsection{Transit timing}
\label{sec:Timing}

\begin{figure}
  \includegraphics[width=1\columnwidth]{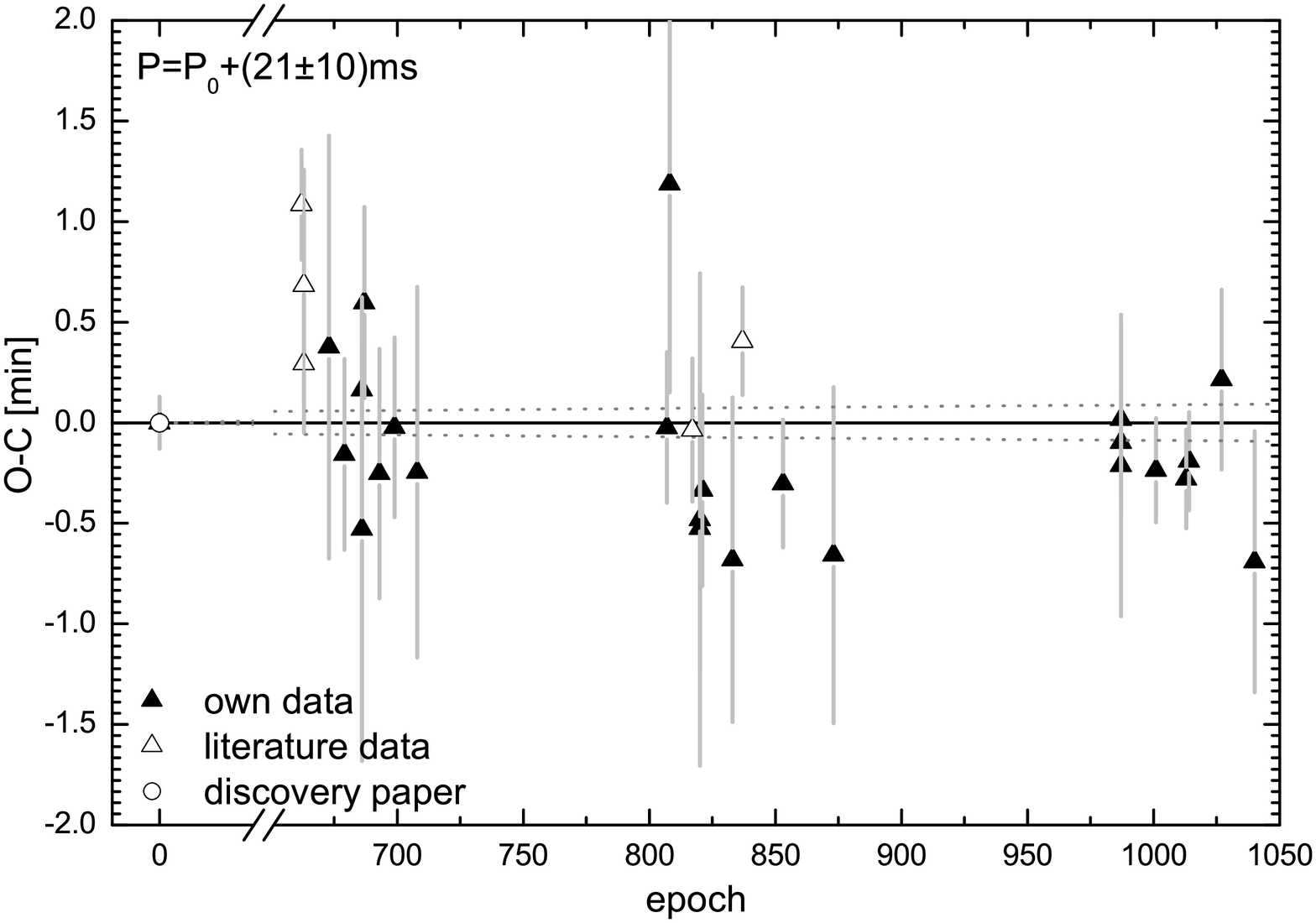}
  \caption{ The O--C diagram for HAT-P-32b assuming the circular orbit parameters from \citet{Hartman2011}. The open circle denotes 
literature data from \citet{Hartman2011}, open triangles denote data from \citet{Sada}, and \citet{Gibson2013}, the filled triangles 
denote our data (from Jena, Tenerife, Rozhen, Sierra Nevada, Swarthmore and Trebur). Redetermining the linear ephemeris by \DeltaP{}
 can explain almost all points (black line denotes the fit, black dotted line the fit error), except for some outliers.}\label{fig:H32OC}
\end{figure} 

The goal of this ongoing project is to look for transit timing variations of known transiting planets where an additional body might 
be present (see section~\ref{sec:Introduction}). In case of HAT-P-32b, we have obtained 24 transit light 
curves with precise mid-transit times out of 45 observations (see Tables~\ref{tab:H32completeObservations} and \ref{tab:H32partialObservations}).
Unfortunately, there are long observational gaps between epochs 720 and 800, and epochs 880 and 980, which is not only due to the 
non-observability in northern summertime, but also due to the bad weather during the last northern winter
 affecting most participating YETI telescopes.

After the fitting process, each obtained mid-transit time has to be converted from JD$_{\textrm{UTC}}$ to barycentric Julian dates in
 the barycentric dynamical time (BJD$_{\textrm{TDB}}$) to account for the Earth's movement. We use the online 
converter\footnotemark\footnotetext{http://astroutils.astronomy.ohio-state.edu/time/utc2bjd.html} made available by Jason 
Eastman \citep*[see][]{Eastman2010} to do the corrections. Since the duration of the transit is just $\sim3\:$ hours, converting
 the mid-transit time is sufficient. The different positions of the Earth at the beginning and the end of the transit event is 
negligible with respect to the overall fitting precision of the mid-transit time, which makes a prior time
 conversion of the whole light curve unnecessary.

As already mentioned in section~\ref{sec:GettingTheLightcurve} it is extremely useful for transit timing analysis to have simultaneous 
observations from different telescopes. Since these data typically are not correlated to each other regarding e.g. the start of observation, 
observing cadence between two images, field of view (and hence number of comparison stars), one can draw conclusions on the 
quality of the data and reveal e.g. synchronization errors. In our case we got simultaneous observations 
at 10 epochs. Unfortunately one of the epoch 673, 693, 807 and 821 observations, and both epoch 834 observations had to be aborted 
due to the weather conditions or technical problems. 
The five remaining simultaneous observations, including the threefold observed transit at epoch 987, are consistent within 
the error bars (see Table~\ref{tab:TimingSimultaneous}). Thus, significant systematic errors can be neglected.

\begin{table}
	\caption{The results of the transit time fits of the four successful simultaneous transit observations. The transit mid-time in epoch 
686, 708, 833, and 987 observations match within the error bars. Epoch 820 observations even match within $3.5\:$s, which is far below the error bars.}
	{\setlength{\tabcolsep}{4pt}
	\label{tab:TimingSimultaneous}
	\begin{tabular}{cll@{$\:\pm\:$}lc}
		\toprule
		epoch &telescope& \multicolumn{2}{c}{$T_{\textrm{mid}}-2\,450\,000\:$d} & $\Delta T_{\textrm{mid}}\:$[d]\\
		\midrule
		686 & Rozhen 2.0m          & $5895.35297$ & $0.00016$ &\multirow{2}{*}{0.00049}\\
		686 & Tenerife 1.2m        & $5895.35248$ & $0.00080$ &\\
		708 & Swarthmore 0.6m      & $5942.65287$ & $0.00064$ &\multirow{2}{*}{0.00108}\\
		708 & Gettysburg 0.4m      & $5942.65179$ & $0.00113$ &\\
		820 & OSN 1.5m             & $6183.45364$ & $0.00085$ &\multirow{2}{*}{0.00003}\\
		820 & Trebur 1.2m          & $6183.45361$ & $0.00049$ &\\
		833 & Trebur 1.2m          & $6211.40361$ & $0.00056$ &\multirow{2}{*}{0.00094}\\
		833 & Jena 0.25m           & $6211.40267$ & $0.00214$ &\\
		987 & Rozhen 2.0m          & $6542.50530$ & $0.00018$ &\multirow{3}{*}{0.00016}\\
		987 & Jena 0.6m            & $6542.50538$ & $0.00032$ &\\
		987 & Torun 0.6m           & $6542.50522$ & $0.00052$ &\\
		\bottomrule
	\end{tabular}
	}
\end{table}

The resultant Observed-minus-Calculated (O--C) diagram is shown in Fig.~\ref{fig:H32OC}. In addition to our data, 
the originally published epoch from \citet{Hartman2011}, three data points from 
\citet{Sada}, and two data points from \citet{Gibson2013} are included.
We can explain almost all points by refining the linear ephemeris by \DeltaPerror. Thus the newly determined period is 

\begin{center}
\begin{tabular}{l@{$\:=\:$}l@{$\:\pm\:$}l}
	$P_{\textrm{new}}$ &(\newP&\newPerror)$\:$d\\
	$P_{\textrm{old}}$ &($2.150008$  &$0.000001\phantom{00}$)$\:$d
\end{tabular}
\end{center}

\begin{figure}
  \includegraphics[width=1\columnwidth]{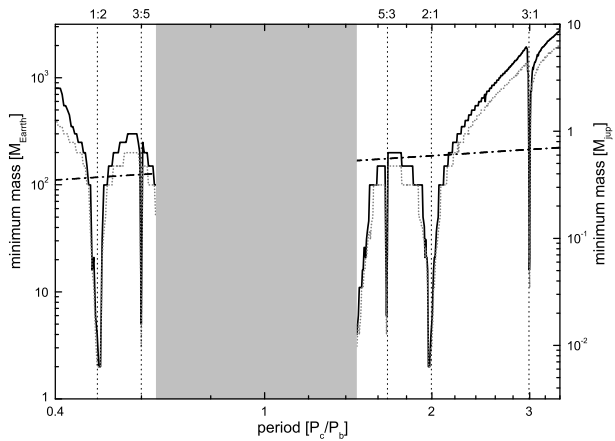}
  \caption{The minimum mass of a perturber needed to create a TTV signal of \Amplitude{} (solid line) and \Amplitudetwo{} (dotted line) 
as a function of the period fraction of the perturber $P_c$ and the known planet $P_b$. \textbf{The dash-dotted line indicates the upper mass limit
for additional planets assuming them to produce a RV amplitude in the order of the observed RV jitter of $\sim63.5\:$m/s.} Within the mean 
motion resonances perturbers with less than $2$ Earth masses can not be excluded. 
The grey shaded area around $P_c/P_b=1$ denotes the dynamically instable region around the known planet 
HAT-P-32b corresponding to $\sim4$ Hill radii.} \label{fig:possibleTTV}
\end{figure}

From the present O--C diagram- we conservatively can rule out TTV amplitudes larger than \Amplitude.
Assuming a circular orbit for both the known planet, and an unseen perturber, we can calculate the minimum perturber mass needed 
to create that signal. In Fig.~\ref{fig:possibleTTV}, all configurations above the solid black line can be ruled out, for they 
would produce a TTV amplitude larger than \Amplitude. Taking a more restrictive signal amplitude of \Amplitudetwo, all masses 
above the dotted line can be ruled out. For our calculations we used the n-body integrator \textit{Mercury6} \citep{Mercury6} 
to calculate the TTV signal for 73 different perturber masses between 1 Earth mass and 9 Jupiter masses placed at 1745 different 
distances to the host star ranging from 0.017$\:$AU to 0.1$\:$AU (i.e. from three times the radius of the host star to three times 
the semimajor axis of HAT-P-32b). These 127385 different configurations have been analysed to search for those systems that produce 
a TTV signal of at least \Amplitude{} and \Amplitudetwo, respectively. In an area around the known planet (i.e. the grey shaded 
area in Fig.~\ref{fig:possibleTTV} corresponding to $\sim4$ Hill radii), most configurations were found to be unstable during the simulated
time-scales. Within the mean motion resonances, especially 
the $1:2$ and $2:1$ resonance, only planets with masses up to a few Earth masses can still produce a signal comparable to (or lower than) 
the spread seen in our data. Such planets would be too small to be found by ground-based observations directly. For distances beyond 
0.1$\:$AU, and accordingly beyond a period ratio above 3, even planets with masses up to a few Jupiter masses would be possible. But 
those planets would generate large, long period transit or radial velocity (RV) signals. Hence one can refuse the existence of such perturbers.

\subsection{Inclination, transit duration and transit depth}
\label{sec:InclinatinoDurationDepth}

\begin{figure}
  \includegraphics[width=1\columnwidth]{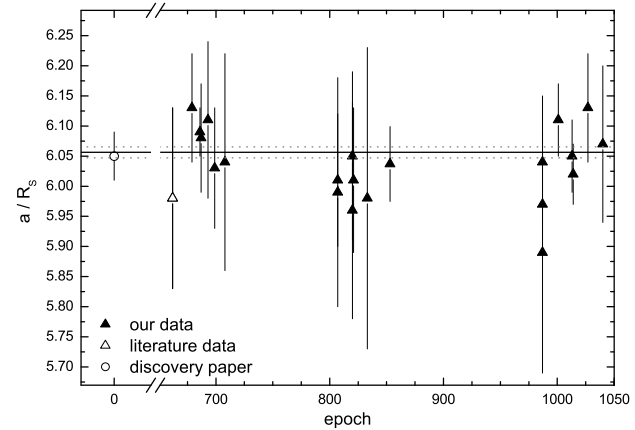}
  \caption{ The obtained values for the ratio of the semimajor axis over the stellar radius $a/R_\textrm{s}$. Assuming a constant value, 
the formal best fit using the model fit errors as instrumental weights is found to be $a/R_\textrm{s}=\,$\newaR$\,\pm\,$\newaRerror$ $ 
(dotted line), which is in agreement with the published value of $a/R_\textrm{s}=6.05\pm0.04$ \citep{Hartman2011}.} \label{fig:H32aR}
\end{figure}

The two transit properties duration and depth are directly connected to the physical properties $a/R_\textrm{s}$ and 
$R_\textrm{p}/R_\textrm{s}$, respectively. Assuming a stable single-planet system, we expect the transit duration to be constant. 
Even if an exo-moon is present, the resultant variations are in the order of a few seconds \citep[for example calculations see][]{Kipping2009} 
and hence too small to be recognized using ground based observations. Using the obtained fitting errors (as described above) as instrumental 
weights for a linear fit, we obtain $a/R_\textrm{s}=\,$\newaR$\,\pm\,$\newaRerror{} (see Fig.~\ref{fig:H32aR}). This value confirms the result 
of \citet{Hartman2011} with $a/R_\textrm{s}=6.05\pm0.04$.

\begin{figure}
  \includegraphics[width=1\columnwidth]{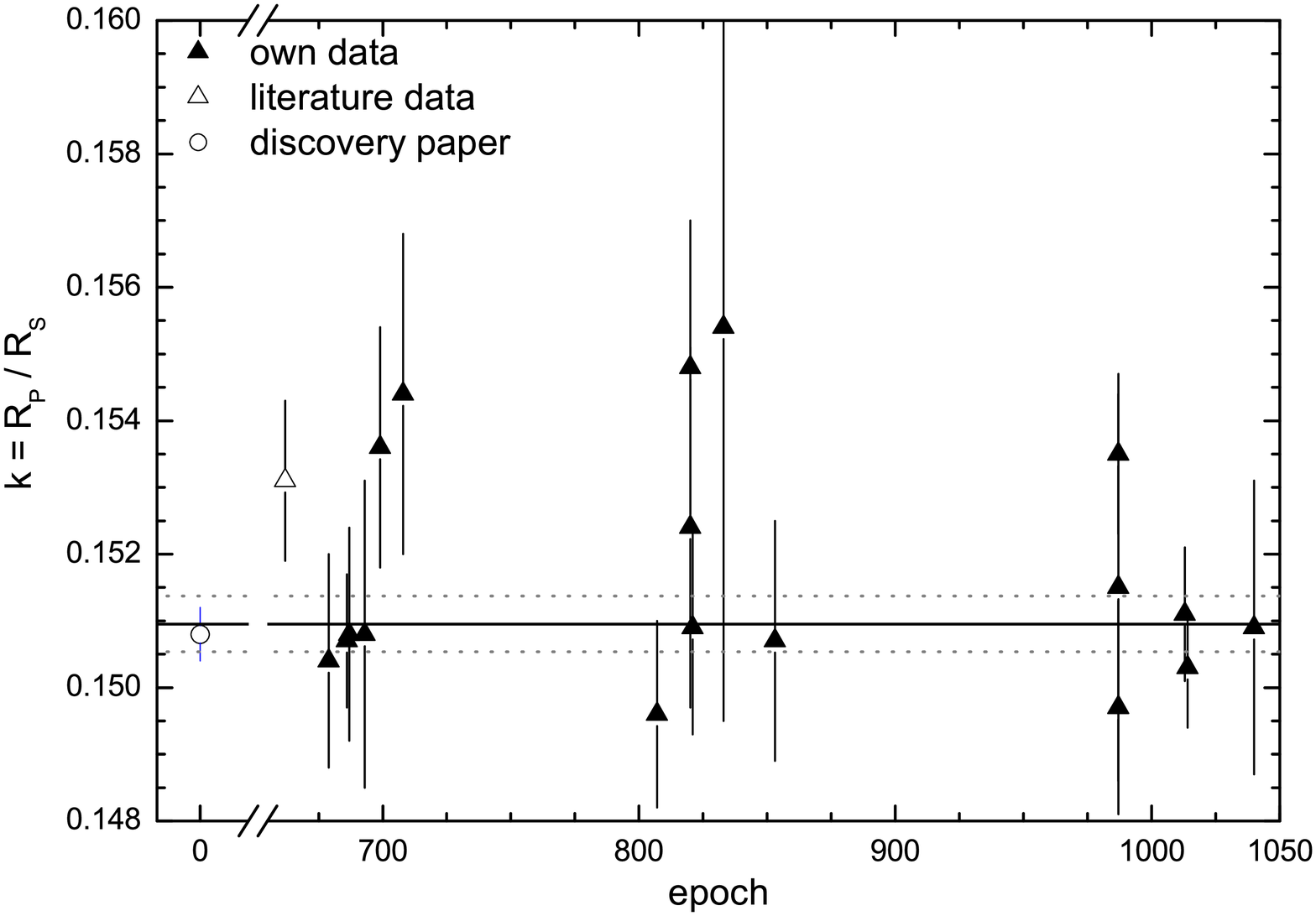}
  \caption{ The obtained values for the ratio of the radii $k=R_\textrm{p}/R_\textrm{s}$. Assuming a constant value, the best fit using 
the model fit errors as instrumental weights is found to be $k=\,$\newk$\,\pm\,$\newkerror$ $ (dotted line), compared to the published 
value of $k=0.1508\pm0.0004$ by \citet{Hartman2011}.}        \label{fig:H32k}
\end{figure}

A similar result can be achieved for the transit depth in Fig.~\ref{fig:H32k} represented by the value of $k=R_\textrm{p}/R_\textrm{s}$. 
Assuming a constant value, we get a fit result of $k=\,$\newk$\,\pm\,$\newkerror{} compared to the originally published value of 
$k=0.1508\pm0.0004$ by \citet{Hartman2011}. 
\citet{Gibson2013} found an M-dwarf $\approx2.8''$ away from HAT-P-32. Though \citet{Knutson2013} ruled out the possibility 
that this star is responsible for the long term trend seen in the RV data, due to the size of our apertures this star 
always contributes to the brightness measurements of the main star and hence can affect the resulting planet-to-star radius ratio. 
Since actual brightness measurements of the star are not available but needed to correct the influence on the parameters, we do not correct 
for the M-dwarf. This way, the results are comparable to those of previous authors, but underestimate the true planet-to-star radius ratio. 
The spread seen in Fig.~\ref{fig:H32k} can therefore be an effect due to the close M-star, or may be also caused by different filter curves 
of different observatories. Possible filter-dependent effects to the transit depth are discussed in Bernt et al. (2014, in prep.).

\begin{figure}
  \includegraphics[width=1\columnwidth]{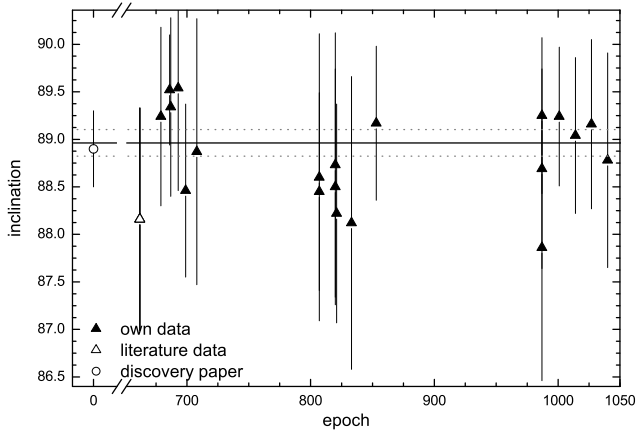}
  \caption{ Inclination versus epoch for the HAT-P-32 system. No change in inclination can be seen. The best fit using the model fit 
errors as instrumental weights is $i=($\newi$\,\pm\,$\newierror$)^\circ$ (dotted line), compared to $\left(88.9\pm0.4\right)^\circ$  
of \citet{Hartman2011}}\label{fig:H32i}
\end{figure} 

As expected, also the inclination is found to be consistent with the originally published value. Moreover, due to the number of data
 points and assuming a constant inclination, we can improve the value to $i=($\newi$\pm$\newierror$)^\circ$ (see Fig~\ref{fig:H32i}). 
Table~\ref{tab:VglHartman} summarizes all our results and compares them to the stellar parameters taken from the circular orbit fit of \citet{Hartman2011}.

\begin{table*}
	\caption{A comparison between the results obtained in our analysis, and the literature data from \citet{Hartman2011}, \citet{Sada}, and \citet{Gibson2013}. All epochs $T_0$ are converted to $BJD_{TDB}$.}\label{tab:VglHartman}
	\begin{tabular}{@{}ll@{$\,\pm\,$}ll@{$\,\pm\,$}ll@{$\,\pm\,$}ll@{$\,\pm\,$}ll@{$\,\pm\,$}l}
	\toprule
				& \multicolumn{2}{c}{$T_0\,[d]$} & \multicolumn{2}{c}{$P\,[d]$} & \multicolumn{2}{c}{$a/R_\textrm{s}$}  & \multicolumn{2}{c}{$k=R_\textrm{p}/R_\textrm{s}$} & \multicolumn{2}{c}{$i\,[^\circ]$} \\
	\midrule
	our analysis		& \newTo	&\newToerror 	 & \newP	&\newPerror  	& \newaR	&\newaRerror		& \newk		&\newkerror & \newi		&\newierror\\
	\citet{Hartman2011}	& $2454420.44637$&$0.00009$	 & $2.150008$	&$0.000001\:$	&\multicolumn{2}{l}{$6.05\phantom{0}\,_{-0.04}^{+0.03}$}& $0.1508$	&$0.0004$& $88.9$		&$0.4$\\
	\citet{Sada}		& $2454420.44637$&$0.00009$	 & $2.1500103$	&$0.0000003$	&\multicolumn{2}{l}{$5.98\phantom{0}\,_{-0.10}^{+0.15}$} & $0.1531$		&$0.0012$& \multicolumn{2}{l}{$86.16\,_{-1.17}^{+1.03}$}\\
	\citet{Gibson2013}	& $2454942.899220$&$0.000077$	 & $2.1500085$	&$0.0000002$ 	& \multicolumn{2}{l}{$6.091\,_{-0.036}^{+0.047}$} & $0.1515$&$0.0012$ &\multicolumn{2}{l}{$89.12\,_{-0.61}^{+0.68}$} \\
	\bottomrule
	\end{tabular}
\end{table*}

\subsection{Further limitations}
\label{sec:furtherLimits}

Besides the transit observations, we are also reanalysing the published RV-data \citep[available at][]{Hartman2011} with
the \textit{systemic console} \citep{Meschiari2009}.
It is indeed possible to increase the precision of the RV fit by putting additional, even lower mass or distant 
bodies into the system. However, due to the large observational gaps seen in Fig.~\ref{fig:H32OC}, the large number of different possible scenarios, 
especially perturbers with larger periods, can hardly be restricted.
Assuming a small perturber mass, an inclination $\sim90^\circ$, and an eccentricity equal to zero, one can easily derive the expected RV amplitude to be
\begin{equation}
	K\simeq\frac{28.4\cdot M_P}{P_c^{1/3}\cdot M_s^{2/3}}\nonumber
\end{equation}
using Keplers laws and the conservation of momentum with the perturber mass $M_P$ in Jupiter masses, its period $P_c$ in days and mass of the central star $M_S$ 
in solar masses. The jitter amplitude of $\sim63.5\:$m/s found by \citet{Knutson2013} then corresponds to a specific maximum mass of a 
potentially third body in the system, depending on its period. Hence, all objects above the dash dotted line in Fig.~\ref{fig:possibleTTV} can be ruled out,
since they would result in even larger RV amplitudes.

In addition to the advantage of simultaneous observations concerning the reliability on the transit timing, one can also use them as quality markers 
for deviations in the light curve itself. As seen in Fig.~\ref{fig:TransitLightCurves3}, there are no systematic differences between the three light curves
obtained simultaneously. This is also true for the data shown in Fig.~\ref{fig:TransitLightCurves2}, where no residual pattern is seen twice, though
there is a small bump in the epoch 686 Teneriffe data. Hence, one can rule out real astrophysical reasons, e.g. the crossing of a spotty 
area on the host star, as cause for the brightness change. In a global context,
no deviations seen in the residuals of our light curves are expected to be of astrophysical origin.

\begin{figure}
  \includegraphics[width=1\columnwidth]{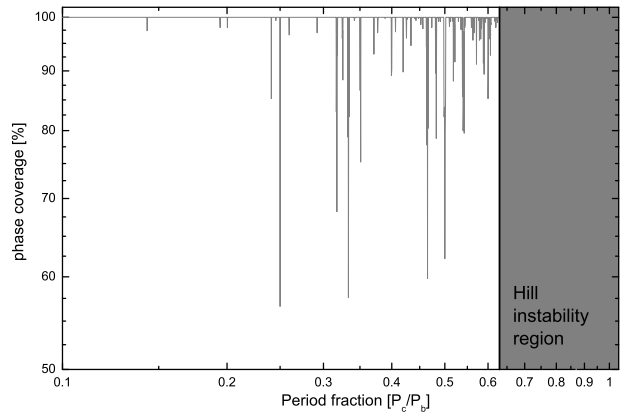}
  \caption{The orbital coverage for an inner transiting perturber as a function of the period fraction of the perturber $P_c$ and the known
transiting planet $P_b$. The unstable area ($\sim4$ Hill radii) around the known transiting planet is marked in grey.}
\label{fig:Coverage}
\end{figure}

Folding the residuals of all light curves obtained within this project to a set of trial periods between $0.15\,$d and $2.15\,$d, 
i.e. period fractions of $P_c/P_b=\{0.1\ldots1\}$, we can analyse the resulting phase folded light curves regarding the orbital coverage.
Thus, we can check if we would have seen the transit of an inner perturber just by chance while observing transits of HAT-P-32b.
Though the duration of a single transit observation is limited to a few hours, taking the large number of observations spread over several months 
we are still able to cover a large percentage of the trial orbits.
As seen in Fig.~\ref{fig:Coverage} we achieve an orbital coverage of more than $90\%$ for the majority of trial periods.
For a small number of certain periods the coverage drops to $\sim80\%$. Especially within the resonances it drops to $\sim60\%$. 
Assuming the detectability of all transit like signatures with amplitudes more than $3\:$mmag (see residuals in Fig.~\ref{fig:TransitLightCurves1}),
we can rule out the existence of any inner planet bigger than $\sim0.5\:$R$_{jup}$. Depending on the composition 
(rocky or gaseous) and bloating status of inner planets, this gives further constraints on the possible perturber mass.

Finally, it is important to state that all findings assume a perturber on a circular orbit, and with the same inclination as the known planet.
It is not unlikely that the inclination of a potential second planet is significantly different from the known transiting planet, as pointed
out e.g. by \citet{Steffen2012} and \citet{PayneFord2011}. This of course also increases the mass needed to create a certain TTV signal. Furthermore, the mass 
range of planets hidden in the RV jitter is also increased. While a possible eccentric orbit of an inner perturber is limited to a small value due to
stability reasons, an outer -- not necessarily transiting -- perturber on an eccentric orbit is also possible, which in turn also would affect the TTV and RV signal. The companion 
candidate discovered by \citet{Gibson2013}, as well as the proposed originator of the RV long term trend \citep{Knutson2013} can, however, not be responsible
neither for the RV jitter, nor for the still present spread in the O--C diagram.

\section{Summary}

\begin{table*}
	\caption{The fit results for the 20 good transit light curves (top rows) as well as the five literature data 
points from \citet{Sada} and \citet{Gibson2013} (middle rows) for the values $T_{\textrm{mid}}$, $a/R_\textrm{s}$, 
$k=R_\textrm{p}/R_\textrm{s}$ and $i$. Since \citet{Sada} did not publish all values for each transit fit, only the
 transit mid time is tabulated for most observations, except the epoch 662 observation, where also the other parameters 
are available. The formal $rms$ and the resultant $pnr$ are given in the last column, if available. In the bottom rows the 
results of the complete observations with larger error bars are given for completeness. In case of the Tenerife 1.2m observations, 
quasi-simultaneous observations in the filters $r_p$ and $B$ have been performed leading to higher $pnr$ values.}
	\label{tab:fitResults}
	\begin{savenotes}
	{\setlength{\tabcolsep}{3.5pt}
	\begin{tabular}{lcll@{$\:\pm\:$}ll@{$\:\pm\:$}ll@{$\:\pm\:$}ll@{$\:\pm\:$}lcc}
		\toprule
		date & epoch &telescope& \multicolumn{2}{c}{$T_{\textrm{mid}}-2\,450\,000\:$d}& \multicolumn{2}{c}{$a/R_\textrm{s}$} & \multicolumn{2}{c}{$k=R_\textrm{p}/R_\textrm{s}$} & \multicolumn{2}{c}{$i\,[^\circ]$} &$rms\:$[mmag] & $pnr\:$[mmag]\\
		\midrule
		2011-11-01 & 673 & Tenerife 1.2m   & $5867.40301$ & $0.00073$ &\multicolumn{2}{c}{--}&\multicolumn{2}{c}{--}&\multicolumn{2}{c}{--}& $2.3$ & $3.20$\\
		2011-11-14 & 679 & Jena 0.6m       & $5880.30267$ & $0.00033$ & $6.13 $ & $0.09 $    & $0.1493$ & $0.0016$  & $89.3 $ & $1.0 $     & $1.9$ & $1.97$\\
		2011-11-29 & 686 & Rozhen 2.0m     & $5895.35297$ & $0.00016$ & $6.09 $ & $0.04 $    & $0.1507$ & $0.0010$  & $89.5 $ & $0.6 $     & $1.0$ & $0.62$\\
		2011-11-29 & 686 & Tenerife 1.2m   & $5895.35249$ & $0.00080$ &\multicolumn{2}{c}{--}&\multicolumn{2}{c}{--}&\multicolumn{2}{c}{--}& $1.7$ & $2.36$\\
		2011-12-01 & 687 & Rozhen 2.0m     & $5897.50328$ & $0.00033$ & $6.08 $ & $0.09 $    & $0.1508$ & $0.0016$  & $89.3 $ & $0.9 $     & $1.5$ & $0.94$\\
		2011-12-14 & 693 & Rozhen 0.6m     & $5910.40274$ & $0.00043$ & $6.11 $ & $0.13 $    & $0.1508$ & $0.0023$  & $89.5 $ & $1.1 $     & $2.0$ & $3.40$\\
		2011-12-27 & 699 & Rozhen 2.0m     & $5923.30295$ & $0.00031$ & $6.03 $ & $0.10 $    & $0.1536$ & $0.0018$  & $88.5 $ & $0.9 $     & $1.2$ & $0.96$\\
		2012-01-15 & 708 & Swarthmore 0.6m & $5942.65287$ & $0.00064$ & $6.04 $ & $0.18 $    & $0.1544$ & $0.0024$  & $88.9 $ & $1.4 $     & $3.3$ & $3.28$\\
		2012-08-15 & 807 & Rozhen 2.0m     & $6155.50385$ & $0.00026$ & $6.01 $ & $0.11 $    & $0.1496$ & $0.0014$  & $88.5 $ & $1.0 $     & $1.3$ & $1.11$\\
		2012-08-18 & 808 & Tenerife 1.2m   & $6157.65470$ & $0.00072$ &\multicolumn{2}{c}{--}&\multicolumn{2}{c}{--}&\multicolumn{2}{c}{--}& $2.6$ & $3.71$\\
		2012-09-12 & 820 & OSN 1.5m        & $6183.45364$ & $0.00085$ & $5.96 $ & $0.18 $    & $0.1524$ & $0.0027$  & $88.7 $ & $1.4 $     & $4.3$ & $4.85$\\
		2012-09-12 & 820 & Trebur 1.2m     & $6183.45361$ & $0.00049$ & $6.05 $ & $0.14 $    & $0.1548$ & $0.0022$  & $88.5 $ & $1.2 $     & $2.1$ & $2.33$\\
		2012-09-14 & 821 & OSN 1.5m        & $6185.60375$ & $0.00033$ & $6.01 $ & $0.12 $    & $0.1509$ & $0.0016$  & $88.2 $ & $1.2 $     & $1.3$ & $1.07$\\
		2012-10-10 & 833 & Trebur 1.2m     & $6211.40361$ & $0.00056$ & $5.98 $ & $0.25 $    & $0.1554$ & $0.0059$  & $88.1 $ & $1.5 $     & $2.2$ & $2.18$\\
		2012-11-22 & 853 & OSN 1.5m        & $6254.40404$ & $0.00022$ & $6.037$ & $0.062$    & $0.1507$ & $0.0018$  & $89.2 $ & $0.8 $     & $1.1$ & $0.84$\\
		2013-01-04 & 873 & Tenerife 1.2m   & $6542.40397$ & $0.00058$ &\multicolumn{2}{c}{--}&\multicolumn{2}{c}{--}&\multicolumn{2}{c}{--}& $1.9$ & $2.87$\\
		2013-09-07 & 987 & Jena 0.6m       & $6542.50538$ & $0.00032$ & $6.04 $ & $0.11 $    & $0.1497$ & $0.0016$  & $88.7 $ & $1.1 $     & $1.6$ & $1.57$\\
		2013-09-07 & 987 & Rozhen 2.0m     & $6542.50530$ & $0.00018$ & $5.97 $ & $0.09 $    & $0.1535$ & $0.0012$  & $88.3 $ & $0.8 $     & $0.9$ & $0.67$\\
		2013-09-07 & 987 & Torun 0.6m      & $6542.50522$ & $0.00052$ & $5.89 $ & $0.20 $    & $0.1515$ & $0.0029$  & $87.9 $ & $1.4 $     & $3.5$ & $2.33$\\
		2013-10-06 &1001 & OSN 1.5m        & $6572.60532$ & $0.00018$ & $6.11 $ & $0.06 $    & $0.1465$ & $0.0013$  & $89.2 $ & $0.7 $     & $0.9$ & $0.63$\\
		2013-11-01 &1013 & Rozhen 2.0m     & $6598.40539$ & $0.00017$ & $6.05 $ & $0.06 $    & $0.1511$ & $0.0010$  & $88.9 $ & $0.8 $     & $0.8$ & $0.68$\\
		2013-11-03 &1014 & OSN 1.5m        & $6600.55546$ & $0.00017$ & $6.02 $ & $0.05 $    & $0.1503$ & $0.0009$  & $89.2 $ & $0.8 $     & $1.3$ & $1.33$\\
		2013-12-01 &1027 & OSN 1.5m        & $6628.50585$ & $0.00031$ & $6.13 $ & $0.09 $    & $0.1475$ & $0.0022$  & $89.2 $ & $0.9 $     & $1.8$ & $1.59$\\
		2013-12-29 &1040 & Trebur 1.2m     & $6656.45533$ & $0.00045$ & $6.07 $ & $0.13 $    & $0.1509$ & $0.0022$  & $88.8 $ & $1.1 $     & $2.6$ & $2.74$\\
		\midrule
		2011-10-09 & 662 & KPNO 2.1m       & $5843.75341$ & $0.00019$ & \multicolumn{2}{l}{$5.98\phantom{0\:}_{-\:0.15}^{+\:0.10}$}&$0.1531$ & $0.0012$&\multicolumn{2}{l}{$88.2\:_{-\:1.0}^{+\:1.2}$} & --&--\\
		2011-10-11 & 663 & KPNO 2.1m       & $5845.90287$ & $0.00024$ &\multicolumn{2}{c}{--}&\multicolumn{2}{c}{--}&\multicolumn{2}{c}{--}& --&--\\
		2011-10-11 & 663 & KPNO 0.5m       & $5845.90314$ & $0.00040$ &\multicolumn{2}{c}{--}&\multicolumn{2}{c}{--}&\multicolumn{2}{c}{--}& --&--\\
		2012-09-06 & 817 &\cite{Gibson2013}& $6177.00392$ & $0.00025$ &\multicolumn{2}{c}{--}&\multicolumn{2}{c}{--}&\multicolumn{2}{c}{--}& --&--\\
		2012-10-19 & 837 &\cite{Gibson2013}& $6220.00440$ & $0.00019$ &\multicolumn{2}{c}{--}&\multicolumn{2}{c}{--}&\multicolumn{2}{c}{--}& --&--\\
		
		\midrule
		2011-10-04 & 660 & Ankara 0.4m     & $5839.45347$ & $0.00101$ & $5.9  $ & $0.2  $    & $0.1448$ & $0.0021$  & $88.1 $ & $1.4 $     & $4.7$ & $2.22$\\
		2012-01-15 & 708 & Gettysburg 0.4m & $5942.65179$ & $0.00113$ & $5.79 $ & $0.36 $    & $0.1493$ & $0.0054$  & $87.3 $ & $1.7 $     & $2.5$ & $4.25$\\
		2012-10-10 & 833 & Jena 0.25m      & $6211.40267$ & $0.00214$ & $6.04 $ & $0.65 $    & $0.1514$ & $0.0089$  & $86.5 $ & $2.5 $     & $5.3$ & $6.96$\\
		2012-10-25 & 840 & Tenerife 1.2m   & $6226.45618$ & $0.00102$ &\multicolumn{2}{c}{--}&\multicolumn{2}{c}{--}&\multicolumn{2}{c}{--}& $3.8$ & $5.05$\\
		2012-12-22 & 867 & Tenerife 1.2m   & $6284.50460$ & $0.00100$ &\multicolumn{2}{c}{--}&\multicolumn{2}{c}{--}&\multicolumn{2}{c}{--}& $3.3$ & $5.10$\\
		\bottomrule
	\end{tabular}}
	\end{savenotes}
\end{table*}

We presented our observations of HAT-P-32b planetary transits obtained during a timespan of 24 months (2011 October until 2013 October). 
The data were collected using telescopes all over the world, mainly throughout the YETI network. Out of 44 started observations we 
obtained 24 light curves that could be used for further analysis. 21 light curves have been obtained that could not be used due to different 
reasons, mostly bad weather. In addition to our data, literature data from \citet{Hartman2011}, \citet{Sada}, and \citet{Gibson2013} were 
also taken into account (see Fig.~\ref{fig:H32OC}, \ref{fig:H32aR}, \ref{fig:H32k}, and \ref{fig:H32i} and Table~\ref{tab:fitResults}).

The published system parameters $a/R_\textrm{s}$, $R_\textrm{p}/R_\textrm{s}$ and $i$ from the circular orbit 
fit of \citet{Hartman2011} were confirmed. In case of the semimajor axis over the stellar radius and the inclination, we were able to 
improve the results due to the number of observations. As for the planet-to-star radius ratio, we did not achieve a better solution for 
there is a spread in the data making constant fits difficult. In addition, \citet{Gibson2013} found an M-dwarf $\approx2.8''$ 
away from HAT-P-32 and hence a possible cause for this spread.

Regarding the transit timing, a redetermination of the planetary ephemeris by \DeltaP{} can explain the obtained mid-transit times, 
although there are still some outliers. Of course, having 1$\sigma$ error bars, one would expect some of the data points to be off the fit. 
Nevertheless, due to the spread of data seen in the O--C diagram, observations are planned to further monitor HAT-P-32b transits using the 
YETI network. This spread in the order of \Amplitude{} does not exclude certain system configurations. Assuming circular 
orbits even an Earth mass perturber in a mean motion resonance could still produce such a signal.

\section*{Acknowledgements}

MS would like to thank the referee for the helpful comments on the paper draft.
All the participating observatories appreciate the logistic and financial support of their institutions and in particular their technical workshops.
MS would also like to thank all participating YETI telescopes for their observations.
MMH, JGS, AP, and RN would like to thank the Deutsche Forschungsgemeinschaft (DFG) for support in the Collaborative Research Center Sonderforschungsbereich SFB~TR~7 ``Gravitationswellenastronomie''.
RE, MK, and RN would like to thank the DFG for support in the Priority Programme SPP 1385 on the \textit{First ten Million years of the Solar System} in projects NE 515/34-1 \& -2.
GM and DP acknowledge the financial support from the Polish Ministry of Science and Higher Education through the luventus Plus grant IP2011 031971.
RN would like to acknowledge financial support from the Thuringian government (B 515-07010) for the STK CCD camera (Jena 0.6m) used in this project.
The research of DD and DK was supported partly by funds of projects DO~02-362, DO~02-85 and DDVU~02/40-2010 of the Bulgarian Scientific Foundation, as well as project RD-08-261 of Shumen University.
We also wish to thank the T\"UB\.{I}TAK National Observatory (TUG) for supporting this work through project number 12BT100-324-0 using the T100 telescope.

\label{lastpage}

\end{document}